\documentclass[graybox]{svmult}

\usepackage{type1cm}        
\usepackage{makeidx}         
\usepackage{graphicx}        
\usepackage{multicol}        
\usepackage[bottom]{footmisc}
\usepackage{bbm}
\usepackage{mathbbol}
\usepackage{bm}
\usepackage{latexsym}
\usepackage{amssymb}
\usepackage{braket}
\usepackage{newtxtext}       %
\usepackage{newtxmath}       
\usepackage{dsfont}

\makeindex             

\newcommand{\bea}{\begin{eqnarray}}
\newcommand{\eea}{\end{eqnarray}}

\newcommand{\avg}[1]{\langle{#1}\rangle}
\newcommand{\Avg}[1]{\left\langle{#1}\right\rangle}

\begin{document}

\title*{Geometry, Topology and Simplicial Synchronization}
\author{Ana Paula Mill\'an, Juan G. Restrepo, Joaqu\'in J. Torres and Ginestra Bianconi}
\institute{Ana Paula Mill\'an \at 
Amsterdam UMC, Vrije Universiteit Amsterdam, Department of Clinical Neurophysiology and MEG Center, Amsterdam Neuroscience, De Boelelaan 1117, Amsterdam, The Netherlands.
\email{a.p.millanvidal@amsterdamumc.nl}
\and Juan G. Restrepo \at Department  of  Applied  Mathematics,  University  of  Colorado  at  Boulder,  Boulder,  CO  80309,  USA\email{juanga@colorado.edu}\and Joaqu{\'i}n J. Torres \at Departamento de Electromagnetismo y Física de la Materia and Instituto Carlos I de Física Teórica y Computacional, Universidad de Granada, 18071, Granada, Spain \email{jtorres@onsager.ugr.es}\and Ginestra Bianconi \at School of Mathematical Sciences, Queen Mary University of London\at The Alan Turing Institute, The British Library, London, UK \email{ginestra.bianconi@gmail.com}}
%
%
\maketitle

\abstract{Simplicial synchronization  reveals  the role that topology and  geometry have in determining the dynamical properties of simplicial complexes.  Simplicial network geometry and topology are naturally encoded in the  spectral properties of the graph Laplacian and of the higher-order Laplacians of simplicial complexes. Here we show how the geometry of simplicial complexes induces  spectral dimensions of the  simplicial complex Laplacians that are responsible for changing the phase diagram of the Kuramoto model. In particular,  simplicial complexes displaying a non-trivial simplicial network geometry cannot  sustain a synchronized state in the infinite network limit if their spectral dimension is smaller or equal to four. This theoretical result is here verified  on the Network Geometry with Flavor simplicial complex generative model displaying emergent hyperbolic geometry. On its turn simplicial topology is shown to  determine the  dynamical properties of the higher-order Kuramoto model. The higher-order Kuramoto model describes synchronization of topological signals, i.e., phases not only associated with the nodes of a simplicial complexes but associated also to higher-order simplices, including links, triangles and so on. This model displays discontinuous synchronization transitions when  topological signals of different dimension and/or their solenoidal and irrotational projections are coupled in an adaptive way.}

\begin{figure}[h]
\sidecaption
\includegraphics[scale=.25]{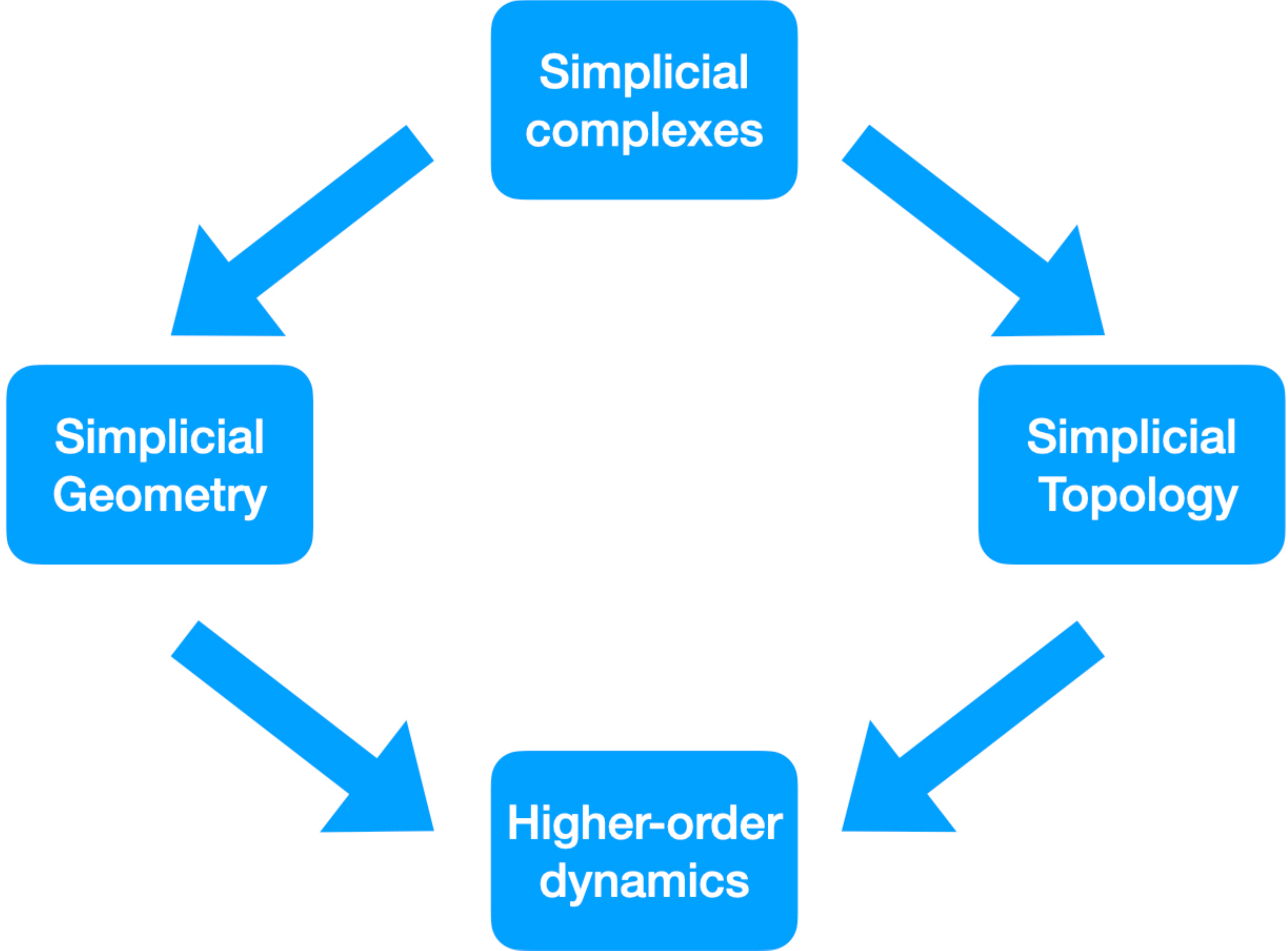}
\caption{\textbf{ Simplicial complexes  encode   the rich simplicial network topology and geometry of data and models, which strongly affects the higher-order dynamics.} In this chapter we will see how this interplay between structure and dynamics can enrich our understanding of synchronization dynamics defined on simplicial complexes.}
\label{fig:1}       
\end{figure}

\section{Introduction}
\label{sec:1}
The interplay between structure and dynamics of complex networks \cite{barabasi2016network,Doro_crit,Barrat} has been at the forefront of network theory since the beginning of the field. In this context it has been found that the combinatorial and statistical  properties of complex networks have unexpected effects on dynamics. For instance, a scale-free degree distribution changes the phase diagram of a wide range of dynamical processes including percolation, epidemic spreading, and the Ising model. The recent  interest on higher-order networks \cite{battiston2020networks,bianconi2021higher,bianconi2015interdisciplinary,eliassirad} provides an opportunity to bring a fresh perspective to this subject. 
Indeed, higher-order networks, and in particular simplicial complexes, constitute the ideal mathematical framework to capture the simplicial network topology and geometry of data. Here we reveal that the network topology and geometry of  simplicial complexes can be crucial to define higher-order dynamics.
The dynamical process considered in this chapter is synchronization \cite{strogatz2000kuramoto,arenas2008synchronization,boccaletti2018synchronization}, captured by the Kuramoto model \cite{kuramoto1975} and the recently introduced higher-order  Kuramoto model \cite{millan2020explosive}.
The dynamical properties of these dynamical processes will be shown to be highly dependent on the spectral properties \cite{torres2020simplicial,Barbarossa} of the simplicial complexes \cite{courtney2016generalized,bianconi2016network,bianconi2017emergent,mulder2018network,courtney2017weighted} on which they are defined. 
The main message of this chapter is summarized in Figure $\ref{fig:1}$, which highlights the role of  network topology and network geometry   in  shaping higher-order network dynamics.
In particular, in this chapter we will disclose  how  the spectral properties of simplicial complexes  are foundational to reveal the relation between higher-order network geometry, topology and dynamics. 
Note that while our approach to simplicial synchronization is based on simplicial network geometry and topology, other approaches based on a combinatorial definition of higher-order interactions have been pursued in the literature \cite{skardal2019abrupt,skardal2019higher,skardal2020memory}, as covered by the Skardal and Arenas chapter of this book.

\begin{figure}[b]
\sidecaption
\includegraphics[scale=.20]{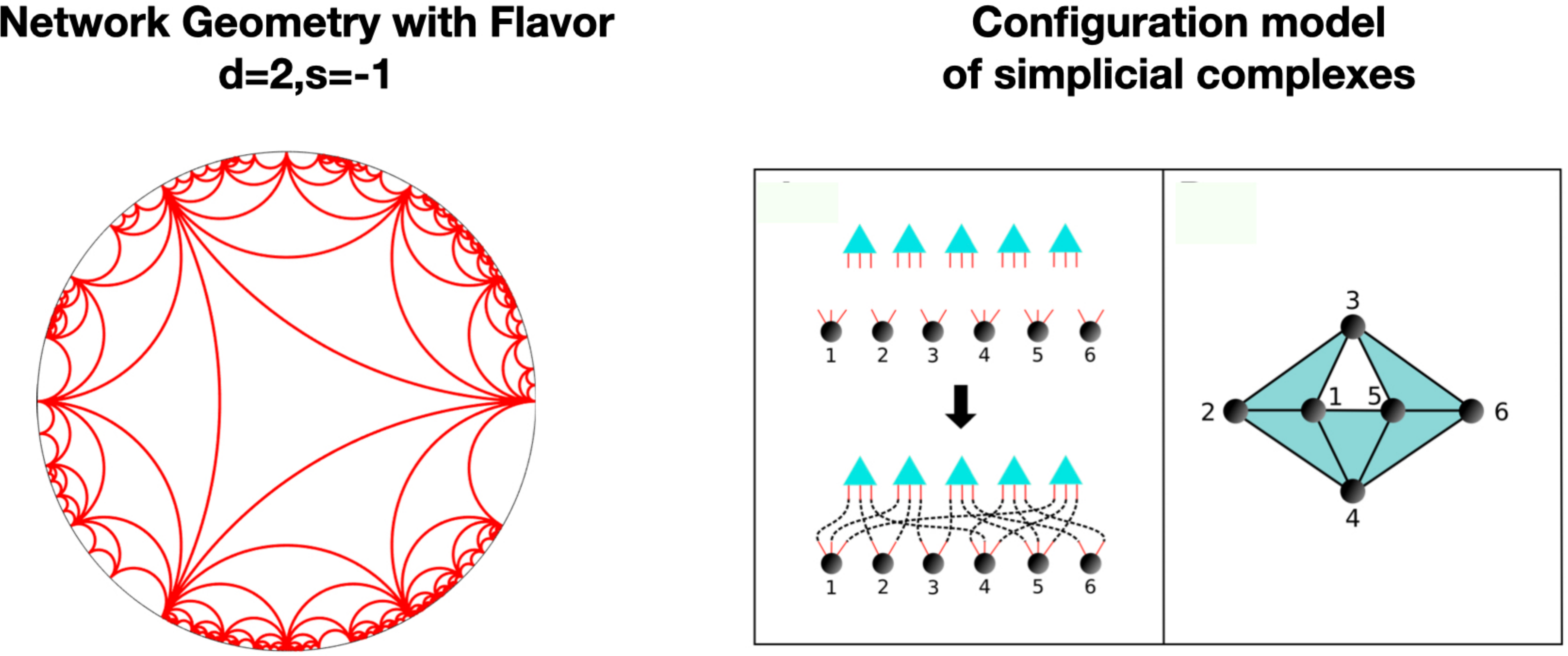}
\caption{ \textbf{Schematic representation of the two classes of simplicial complex models considered in this work.} 
The Network Geometry with Flavor is a model of growing simplicial complexes describing emergent hyperbolic network geometries. The left panel shows a realization of the  NGF simplicial complex of dimension $d=2$ and flavor $s=-1$. The configuration model  of simplicial complexes (right panel) is a maximum entropy model enforcing a given sequence of generalized degrees of the nodes.  Right panel  reprinted    with  permission  from  Ref.~\cite{courtney2016generalized}.  \copyright Copyright  (2016)  by  the American Physical Society.}
\label{fig:2}      
\end{figure}

\section{Simplicial complex models}
\label{sec:2}
Simplicial network models are ideal to test, in a well-controlled setting, the interplay between simplicial network geometry, topology and dynamics. 
Here we focus in particular on two large classes of simplicial complex models with very distinct structural properties: the Network Geometry with Flavor (NGF) \cite{bianconi2016network,bianconi2017emergent,mulder2018network,courtney2017weighted}  and the configuration model of simplical complexes \cite{courtney2016generalized} (see schematic illustrations of the two models in Figure $\ref{fig:2}$). These models are implemented in codes available at the repository \cite{GitHubrepository}.

\subsection{The Network Geometry with Flavor (NGF)}

The Network Geometry with Flavor (NGF) \cite{bianconi2016network,bianconi2017emergent,mulder2018network,courtney2017weighted}      is a general mathematical framework for growing simplicial complexes that displays emergent hyperbolic network geometry. In other words, the NGF model generates simplicial complexes with hyperbolic geometry  that evolve following purely combinatorial and stochastic rules that do not make any use of the natural hyperbolic embedding  of the simplicial complexes. 

The NGFs are simplicial complexes characterized by two main parameters: the dimension of the simplicial complex $d$ and the {\em flavor} $s$ which is a parameter that takes values $s\in \{-1,0,1\}$.
The NGFs are generated by a dynamical process which starting at time $t=1$ from a single $d$-dimensional simplex proceeds at each time $t>1$ by adding a new $d$-dimensional simplex to the simplicial complex. The new $d$-dimensional simplex includes one new node and is glued  to a $(d-1)$-dimensional face $\alpha$ of the existing simplicial complex chosen according to the probability 
\bea
\Pi_{\alpha}=\frac{1-s+sk_{d,d-1}(\alpha)}{\sum_{\alpha^{\prime}}1-s+sk_{d,d-1}(\alpha^{\prime})},
\eea 
where $k_{d,d-1}(\alpha)$ indicates the generalized degree \cite{courtney2016generalized} of a $(d-1)$-dimensional face $\alpha$, i.e., the number of $d$-dimensional simplices incident to the $(d-1)$-dimensional face $\alpha$.
This model generates emergent hyperbolic simplicial complexes which satisfy Gromov criteria \cite{Gromov} of hyperbolicity and are $\delta$-hyperbolic with $\delta=1$ for every value of the flavor $s$ \cite{millan2021local}. 
Moreover, for flavor $s=-1$ the generated simplicial complexes form $d$-dimensional hyperbolic  manifolds \cite{bianconi2017emergent}. The network skeleton of the NGFs are small world, display hierarchical community structure and are  scale-free for $d\geq 2-s$ \cite{bianconi2016network,bianconi2017emergent,mulder2018network}.
 Interestingly, in the case  $d=1$ and $s=1$ the NGF reduces to the Barab\'asi-Albert model, and for $d=1$ and $s=-1$ the NGF reduces to random Apollonian networks.

This model can be generalized in different directions. Instead of considering simplicial complexes, one can use a similar model to generate cell complexes by gluing together convex regular polytopes \cite{mulder2018network}. Another possibility is to consider weighted simplicial complexes or to allow any new node to be incident to  more than one $d$-dimensional simplex \cite{courtney2017weighted}. Finally, the faces can be assigned a fitness that can be used to modulate the attachment probability $\Pi_{\alpha}$  causing topological phase transitions for certain fitness distributions  \cite{bianconi2016network,bianconi2017emergent}.

\subsection{Configuration model of simplicial complexes}
The configuration model of simplicial complexes \cite{courtney2016generalized} is a maximum entropy model of pure $d$-dimensional simplicial complexes.
A pure $d$-dimensional simplicial complex $\mathcal{K}$ can be fully encoded in a $(d+1)$-dimensional adjacency tensor indicating the presence of each $d$-dimensional facet of the simplicial complex.
In particular adjacency tensor ${\bf a}$ has elements $a_{\alpha}=1$ if the $d$-dimensional simplex $\alpha$ is present in the simplicial complex, otherwise $a_{\alpha}=0$.

The configuration model of simplicial complexes \cite{courtney2016generalized} is the least biased ensemble of simplicial complexes 
 with a given generalized degree sequence of the nodes  \bea
{\bf k}_{d,0}=\{k_{d,0}(1),k_{d,0}(2),\ldots k_{d,0}(N)\}
\eea
where $k_{d,0}(i)$ indicates the generalized degree of the node $i$, i.e., the number of $d$-dimensional complexes of the node $i\in\{1,2,\ldots, N\}$.

The configuration  model of simplicial complexes is fully characterized by the probability $P(\mathcal{K})$ assigned to each pure $d$-dimensional simplicial complex $\mathcal{K}$ of $N$ nodes. The probability $P(\mathcal{K})$ maximizes the entropy $S$ of the simplicial complex ensemble 
\bea
S=-\sum_{\mathcal{K}}P(\mathcal{K})\ln P(\mathcal{K}),
\eea
given the constrain that each node $i$ has generalized degree $k_{d,0}(i)$, i.e.,
\bea
\sum_{\alpha\in Q_d(N)|i\subset \alpha}a_{\alpha}=k_{d,0}(i),
\eea
where $Q_d(N)$ indicates the set of all possible $d$-dimensional simplices of a simplicial complex formed by $N$ nodes.
Therefore the configuration model of simplicial complexes is characterized by the uniform distribution 
\bea
P(\mathcal{K})=\frac{1}{\mathcal{N}}\prod_{i=1}^N\delta\left(\sum_{\alpha\in Q_d(N)|i\subset \alpha}a_{\alpha},k_{d,0}(i)\right),
\eea
where here and in the following $\delta(a,b)$ indicates the Kronecker delta.
In Ref. \cite{courtney2016generalized} the Authors proposed an algorithm for sampling simplicial complexes from this ensemble. This algorithm \cite{GitHubrepository} uses the mapping of simplicial complexes to factor graphs. 
The configuration model of simplicial complexes is a very valuable null model of simplicial complexes, and provides an ideal benchmark to study dynamical processes on higher-order networks.

\section{Laplacians}
\subsection{Graph Laplacian}

The graph Laplacian describes linear diffusion on a network and it is an important operator that encodes the interplay between network structure and dynamics \cite{Barrat,burioni2005random}. The graph Laplacian can also capture the underlying network geometry of the skeleton of a simplicial complex, i.e., of the network obtained from the simplicial complex by retaining only its nodes and links.

The graph Laplacian $\bm{L}_{[0]}$ is the discrete version of the Laplacian operator defined on continuous functions. The graph Laplacian of a network with $N_{[0]}$ nodes is a $N_{[0]}\times N_{[0]}$ matrix of elements  
\begin{equation}\label{eq:laplacian_1d}
\left[L_{[0]}\right]_{ij} = k_i\delta({i,j}) - a_{ij},
\end{equation} 
where $k_i$ is the degree of node $i$ and $a_{ij}$ are the elements of the adjacency matrix of the network.
In some cases it is useful to consider a generalization of this operator called the  normalized Laplacian $\bm{\hat{L}}_{[0]}$ that has elements
\begin{equation}\label{eq:laplacian_normalized_1d}
\left[\hat{L}_{[0]}\right]_{ij} = \delta({i,j}) - \frac{a_{ij}}{k_i}.
\end{equation}
For instance, the normalized Laplacian is  commonly employed to describe random walk dynamics on a network.

Both the standard and the normalized Laplacians have real eigenvalues $0=\lambda_1 \leq \lambda_2 \leq\ ...\ \leq \lambda_{N_{[0]}}$. The density of eigenvalues is described by the spectral density, 
\begin{equation}\label{eq:rho_1d}
    \rho(\lambda) = \frac{1}{N_{[0]}} \sum_{i=1}^{N_{[0]}} \tilde{\delta}(\lambda-\lambda_i),
\end{equation}
where $\tilde\delta\left(x\right)$ indicates the delta function. 
The eigenvectors of the normalized Laplacian also encode relevant properties of the underlying network.

\subsection{Spectral dimension}
 
For many complex networks, the smallest non-zero eigenvalue (also called Fiedler eigenvalue) $\lambda_2$ remains finite as the network size increases.  In this case, the network is said to have a \emph{spectral gap}.
On the contrary, if $\lambda_2\to 0$ as $N\to \infty$, and the density of eigenvalues $\rho(\lambda)$ scales as
\begin{equation}\label{eq:rho_1d}
    \rho(\lambda) \propto \lambda^{d_S^{[0]}/2-1},
\end{equation}
for $\lambda\ll1$, the network is said to present a  \emph{spectral dimension} $d_S^{[0]}$ \cite{rammal1983random,burioni2005random,burioni1996universal}.
The spectral dimension can be interpreted as the perceived dimension of the network by diffusion processes, and it is a notable feature of networks with an underlying geometrical nature. 
This is a definition of dimension that is alternative to the Hausdorff  dimension $d_H$ characterizing the scaling of the diameter $D$ of a network with the network size $N$, i.e., $D\propto N^{1/d_H}$.

For Euclidean lattices of dimension $d$, the spectral dimension coincides with the Hausdorff dimension and we have  $d_S^{[0]}=d_H=d$. However, in general the spectral dimension of a network skeleton of a simplicial complex does not have to coincide with the topological dimension of the simplicial complex, nor with its Hausdorff dimension \cite{wu2015emergent,rammal1983random, millan2020complex,millan2021local}.

While Euclidean lattices display a finite spectral dimension,  random graphs and the configuration model of networks are instead characterized by a finite spectral gap. The presence of a finite spectral gap indicates the mean-field nature of the network interactions, and the absence of a clear notion of locality for these networks. 
Interestingly, also the configuration model of simplicial complexes displays a finite spectral gap. 

Remarkably, the emergent network geometry of NGF reveals itself on their significant spectral properties. Indeed the NGF network skeletons, together with other small-world of simplicial complexes \cite{tadic1}, have a finite spectral dimension whose value can be tuned according to the different control parameters \cite{mulder2018network} although the NGFs are small world, i.e., they have an infinite Hausdorff dimension $d_H=\infty$.
 \begin{figure}
    \centering
    \includegraphics[width=0.8\textwidth]{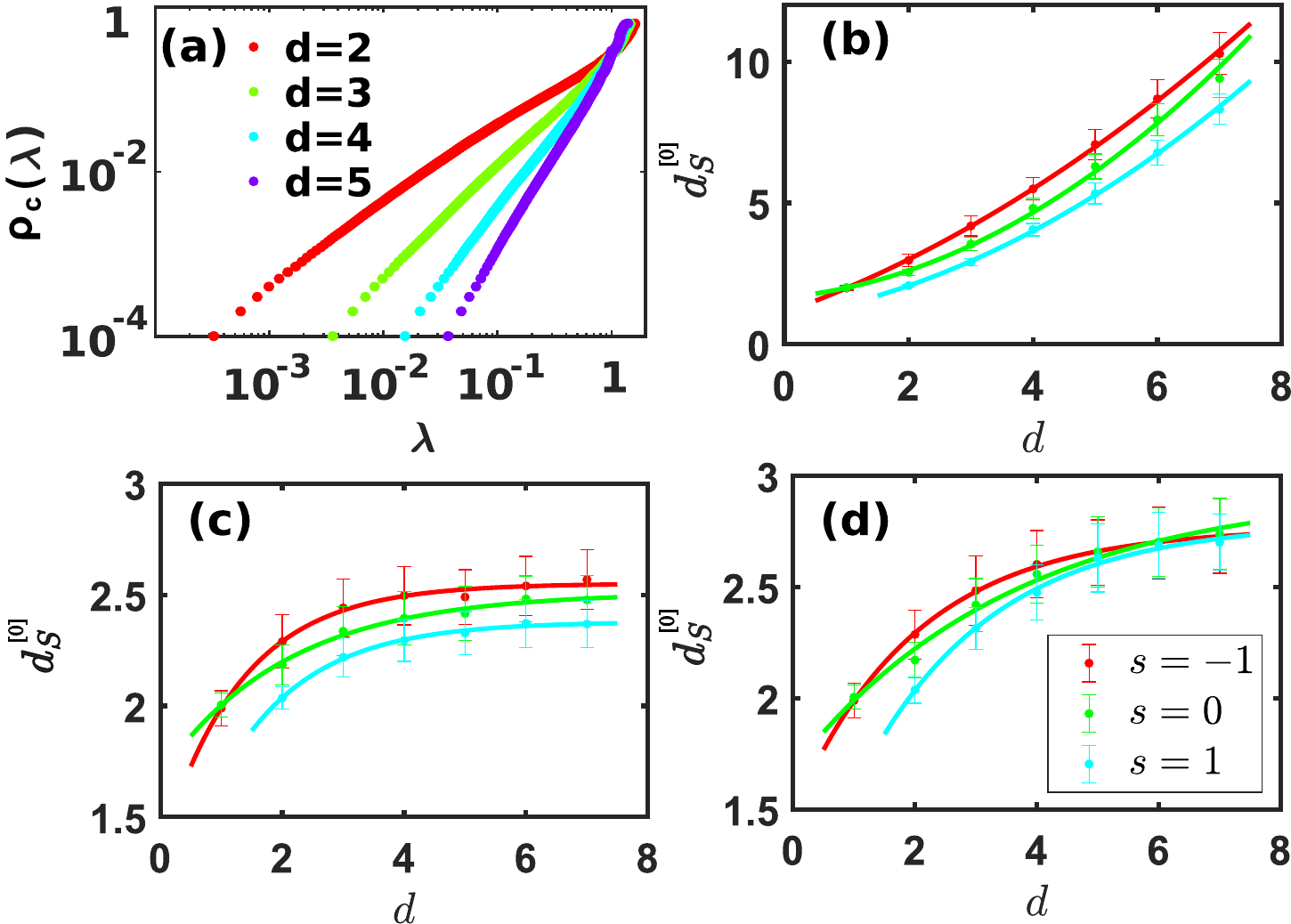}
    \caption{\textbf{Spectral dimension of NGF networks.}  \textbf{Panel (a)} The cumulative distribution $\rho_c(\lambda)$ of eigenvalues of the  NGF with flavor $s=-1$ is shown for  dimension $d=2,3,4,5$. 
  The power-law scaling of   $\rho_c(\lambda)$  observed for small values of $\lambda$ indicates that the skeleton of NGF has a finite spectral dimension. 
    \textbf{Panels (b-d)} The fitted spectral dimension $d_S^{[0]}$ of the skeleton of NGF simplicial and cell complexes being formed by  simplices (panel (\textbf{b})), hypercubes (panel (\textbf{c})) and orthoplexes (panel (\textbf{c})), is shown for values of the  flavor $s\in\{-1,0,1\}$ as indicated in the legend of panel (\textbf{d}). 
    Lines indicate best fit of the $d_S^{[0]}$ versus $d$ dependence using  parabolic (panel (\textbf{b})) and exponential (panels (\textbf{c}), (\textbf{d})) functional forms. Data from Ref. \cite{millan2019synchronization} and Ref. \cite{mulder2018network}. Details of the fits are described in Ref. \cite{mulder2018network}.  }
    \label{fig:s3_fig1}.
\end{figure}
For $s=-1$, NGF networks formed purely by $d$-dimensional simplices have $d_S^{[0]}\sim d$ for $d\in\{2,3,4\}$
as shown in Figure $\ref{fig:s3_fig1}$ \cite{millan2018complex}.
More generally, for NGFs formed by regular polytopes, $d_S^{[0]}$ increases with the dimension $d$ of the polytopes, and it saturates for hypercubes and orthoplexes ($d_S^{[0]}\leq 3$) \cite{millan2019synchronization}. 
It was shown in Ref. \cite{mulder2018network} that a similar trend is observed for different flavours of the NGF networks: $d_S^{[0]}$ grows proportional to $d^2$ for simplicial NGF networks, whereas it saturates at a value $d_S^{[0]}=\bar{d}_S$ with $2 \leq \bar{d}_S\leq 3$ for hypercube and orthoplex NGF networks. 
Moreover, it was shown recently in Ref. \cite{millan2021local} that, in a generalization of the NGF model in which different polytopes are glued together in the same higher-order network, the spectral dimension of the network skeleton can be continuously tuned as a function of the fraction of simplexes in the cell complex. 

Thus, not only the dimension of the building blocks shapes the spectral dimension of the networks, but the specific nature and symmetry of these building blocks also play a role in the emerging spectral dimension of the network skeleton (see Figure $\ref{fig:s3_fig1}$).

\subsection{Higher-order Laplacians}

Important topological aspects of simplicial complexes are reflected in the spectral properties of the higher-order Laplacians that generalize the graph Laplacian to describe diffusion that occurs among higher-order simplices.
The graph Laplacian describes diffusion occurring among nodes connected by links. Similarly the $n$-th order up-Laplacian describes the diffusion occurring among  $n$-dimensional simplices connected by $(n+1)$-dimensional simplices and the  $n$-th order down-Laplacian describes the diffusion occurring among $n$-dimensional simplices connected by $(n-1)$-dimensional simplices.
The higher-order Laplacians capture the topology of simplicial complexes. For instance their spectrum encodes the Betti numbers, i.e., the number of $n$-dimensional cavities of the simplicial complex.
 
The higher-order Laplacians are defined in terms of the incidence matrices of the simplicial complex which represent the boundary operators playing a fundamental role in algebraic topology.

Here we provide a brief introduction to the necessary elements of algebraic topology needed to define higher-order Laplacians.

\begin{figure}[h]
    \centering
    \includegraphics[width=0.8\textwidth]{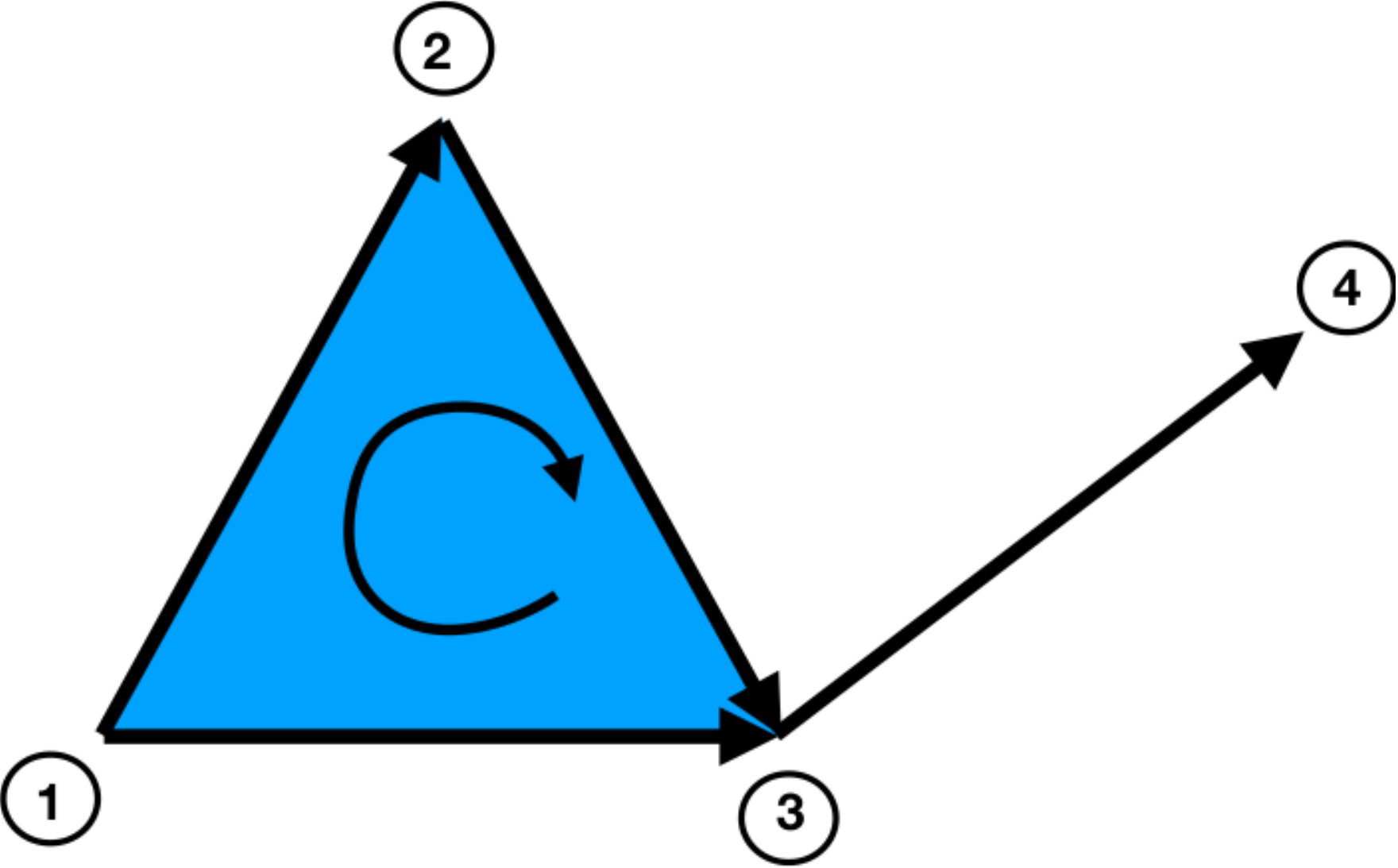}
    \caption{\textbf{Example of oriented simplicial complex.} This illustrates an example of a $2$-dimensional oriented simplicial complex having associated incidence matrices given by Eqs. (\ref{incidence_example}). Reprinted figure with permission from Ref. \cite{millan2020explosive}. \copyright   Copyright  (2020)  by  the American Physical Society.}
    \label{fig:4}.
\end{figure}

We consider a $d$-dimensional simplicial complex formed by $N_{[n]}$ simplices of dimension $n$. The simplices of the simplicial complexes are associated with an orientation induced by the labelling of the nodes so that the link $[i,j]$ has a positive orientation if $j>i$ and so on (see Figure $\ref{fig:4}$).

We consider algebraic entities called $n$-chains that are  linear combinations of $n$-dimensional  simplices with coefficients in $\mathbb{Z}$.
In a less informal definition $n$-chains  are the elements of a free abelian group ${\mathcal C}_n$ with basis on the $n$-simplices of the simplicial complex. The boundary map is a  linear map $\partial_n:{\mathcal C}_n\to {\mathcal C}_{n-1}$ defined by its action on each simplex $\alpha=[i_0,i_1,i_2\ldots,i_n]$. In particular  the boundary map  associates  to every  $n$-dimensional simplex $\alpha=[i_0,i_1,i_2\ldots,i_n]$  a linear combination of the $(n-1)$-dimensional oriented faces at its boundary, given by 
\bea
\partial_n [i_0,i_1\ldots,i_n]=\sum_{p=0}^n(-1)^p[i_0,i_1,\dots,i_{p-1},i_{p+1},\dots,i_n].
\label{boundary}
\eea
It follows that the image of the boundary operator $\partial_n$ are the $(n-1)$-chains that are at the boundary of $n$-chains.
For instance we have 
\bea
\partial_{2}([1,2,3])=[1,2]+[2,3]-[1,3],
\eea
i.e., the image of a triangle is the linear combination of the links at its boundary with the correct orientation.
Additionally, from this definition it is also easy to see that a cyclic $n$-chain is in the  kernel of the boundary map  $\partial_n$ 
independently of whether the cyclic $n$-chain is the boundary of a $(n+1)$-chain.
For instance we have 
\bea
\partial_{1}([1,2]+[2,3]-[1,3])=[2]-[1]+[3]-[2]-[3]+[1]=0,
\eea
whether the simplex $[123]$ belongs to the simplicial complex or not.
One important topological property of the boundary operator is that the ``boundary of a boundary is null" which implies 
\bea
\mbox{im} \  \partial_{n+1}\subseteq \mbox{ker}  \ \partial_{n}
\eea
or equivalently 
\bea
\partial_n\partial_{n+1}=0.
\eea
For instance we have 
\bea
\partial_{1}\partial_2[1,2,3]=\partial_{1}([1,2]+[2,3]-[1,3])=0.
\eea
The boundary map $\partial_n$ can be represented by a  $N_{[n-1]}\times N_{[n]}$ incidence matrix ${\bf B}_{[n]}$ if we adopt
as a basis of the space ${\mathcal C}_n$ an ordered set of the $n$-dimensional simplices $\alpha $, and as a basis of the space ${\mathcal C}_{n-1}$ an ordered set of the $(n-1)$-dimensional simplices $\hat{\alpha} $.

If the  basis of $n$-chains ${\mathcal C}_{n}$ is given by  the $n$-simplices $\{\alpha_1,\alpha_2,\ldots\alpha_s \ldots\}$ and the basis of $(n-1)$-chains  ${\mathcal C}_{n-1}$ is given by the  $(n-1)$-simplices $\{\hat{\alpha}_1,\hat{\alpha}_2,\ldots \hat{\alpha}_r \ldots\}$ the action of the boundary map over any arbitrary $n$-dimensional simplex $ \alpha_s=[i_0,i_1\ldots,i_n]$ given by Eq. ($\ref{boundary}$) can be  expressed as  
\bea
\partial_n \alpha_s=\sum_{r=1}^{N_{n-1}} [B_{[n]}]_{rs} \hat{\alpha}_r.
\label{B}
\eea
This equation  fully determines the incidence matrices ${\bf B}_{[n]}$.  Since we have seen that the ``boundary of a boundary is null" then the incidence matrices follow  ${\bf B}_{[n]}{\bf B}_{[n+1]}={\bf 0}$ and also ${\bf B}_{[n+1]}^{\top}{\bf B}_{[n]}^{\top}={\bf 0}$.
As an example we can consider the simplicial complex shown in Figure $\ref{fig:4}$ whose incidence matrices are given by  
\bea
{\bf B}_{[1]}=\begin{array}{c|cccc}
&$[1,2]$&$[1,3]$&$[2,3]$&$[3,4]$\\
\hline
$[1]$&-1&-1 &0&0\\
$[2]$&1&0&-1&0\\
$[3]$&0&1&1&-1\\
$[4]$&0&0&0&1\\
\end{array},
\ 
{\bf B}_{[2]}=\begin{array}{c|c}
 &$[1,2,3]$\\
 \hline
$[1,2]$&1\\
$[1,3]$&-1\\
$[2,3]$&1\\
$[3,4]$&0
\end{array}.
\label{incidence_example}
\eea

The graph Laplacian can be   expressed in terms of the incidence matrix  ${\bf B}_{[1]}$  of the graph as 
\bea
{\bf L}_{[0]}={\bf B}_{[1]}{\bf B}^{\top}_{[1]}.
\eea 
Similarly, in a simplicial complex the higher-order Laplacian ${\bf L}_{[n]}$ (with $n>0$) \cite{simplices2,thesis,Barbarossa} is the  $N_{[n]}\times N_{[n]}$ matrix defined as 
\bea
    {\bf L}_{[n]}={\bf L}^{[\text{down}]}_{[n]}+{\bf L}^{[\text{up}]}_{[n]},
    \label{hlap}
\eea
where 
\bea
{\bf L}^{[\text{down}]}_{[n]}={\bf B}^{\top}_{[n]}{\bf B}_{[n]},\quad{\bf L}^{[\text{up}]}_{[n]}={\bf B}_{[n+1]}{\bf B}^{\top}_{[n+1]}.
\eea

The $n$-Laplacian is positive semi-definite and, therefore, it has $N_{[n]}$ non negative eigenvalues
$0\leq \lambda_1\leq \lambda_2\leq \ldots \lambda_r\leq \ldots \leq \lambda_{N_{[n]}}.$ A notable property of the spectrum of the $n$-th order  Laplacian is that the degeneracy of zero eigenvalues is given by the $n$-th Betti number. 
Despite the fact that the construction of the  higher-order Laplacians described above seems to rely on the choice of the orientations adopted for the simplices of the simplicial complex, it is possible to show that the higher-order Laplacians are independent on the orientation of the simplices as long as such orientation is induced by the labeling of the nodes.

We note that also the up and the down Laplacians are  positive semi-definite and that from the definition of these matrices it follows immediately than the non-zero eigenvalues in the spectrum of ${\bf L}_{[n]}^{[\text{up}]}$ are the same as the non-zero eigenvalues in the  spectrum of ${\bf L}_{[n]}^{[\text{down}]}$.

By considering the property of the incidence matrices such that ${\bf B}_{[n]}{\bf B}_{[n+1]}=0,$ it is possible to derive the Hodge decomposition of the space of $n$-chains, which reads
\bea
\mathbb{R}^{D_n}=\mbox{img}({\bf B}_{[n]}^{\top}) \oplus \mbox{ker}({\bf L}_{[n]})\oplus \mbox{img}({\bf B}_{[n+1]}).
\eea
This implies that the higher-order Laplacian ${\bf L}_{[n]}$ can be simultaneously diagonalized with the $n$-th order up and $n$-th order down Laplacian and that the non-zero eigenvectors of ${\bf L}_{[n]}$ are either non-zero eigenvector of ${\bf L}_{[n]}^{[\text{down}]}$ or non-zero eigenvectors of ${\bf L}_{[n]}^{[\text{up}]}$. Therefore there is a basis in which ${\bf L}_{[n]}$, ${\bf L}_{[n]}^{[\text{down}]}$ and ${\bf L}_{[n]}^{[\text{up}]}$ have diagonal form given by 
\bea
{\bf U}^{-1}{\bf L}_{[n]}{\bf U}=\left(\begin{matrix}{\bf D}^{[\text{down}]}_{[n]} & {\bf 0} & {\bf 0}\\
{\bf 0} & {\bf 0} & {\bf 0}\\
{\bf 0} & {\bf 0} & {\bf D}^{[\text{up}]}_{[n]}\end{matrix}\right), \ {\bf U}^{-1}{\bf L}_{[n]}^{[\text{down}]}{\bf U}=\left(\begin{matrix}{\bf D}^{[\text{down}]}_{[n]} & {\bf 0} & {\bf 0}\\
{\bf 0} & {\bf 0} & {\bf 0}\\
{\bf 0} & {\bf 0} & {\bf 0} \end{matrix}\right), \ {\bf U}^{-1}{\bf L}_{[n]}^{[\text{up}]}{\bf U}=\left(\begin{matrix} {\bf 0} & {\bf 0} & {\bf 0}\\
{\bf 0} & {\bf 0} & {\bf 0}\\
{\bf 0} & {\bf 0} & {\bf D}^{[\text{up}]}_{[n]}\end{matrix}\right),\nonumber 
\eea
where ${\bf D}_{[n]}^{[\text{up}]}$ and ${\bf D}_{[n]}^{[\text{down}]}$ are diagonal matrices having positive diagonal elements.
\subsection{Higher order spectral dimension}

The  notion of spectral dimension (see section 3.2) can be generalized to  $n$-order up-Laplacians with important consequences for higher-order simplicial complex dynamics. Here, we will focus on the model of NGF simplicial complexes introduced in section 2.1. 
In section 3.2, we have shown that the  graph Laplacian of NGFs  displays a finite spectral dimension $d_S^{[0]}$  \cite{mulder2018network, millan2018complex, millan2019synchronization}. 
Interestingly, higher-order up-Laplacians ${\bf L}_{[n]}^{[\text{up}]}$ and the higher-order down-Laplacians ${\bf L}_{[n]}^{[\text{down}]}$ of NGFs also display a finite spectral dimension.

 \begin{figure}
\begin{center}
\includegraphics[width=\textwidth]{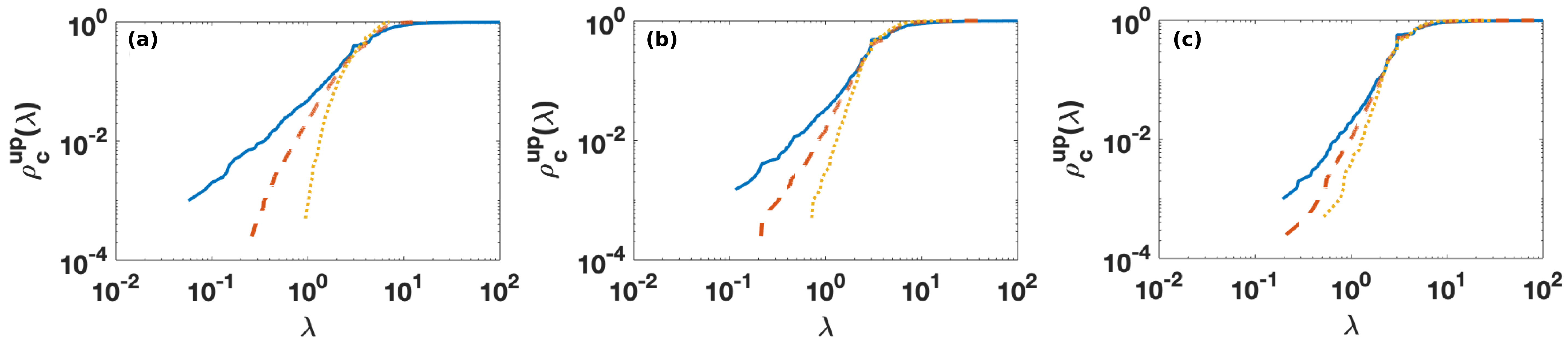}
\end{center}
\caption{  The cumulative density    of non-zero eigenvalues $\rho_c^{up}(\lambda)$ of the ${\bf L}^{[\text{up}]}_{[n]}$  for NGF  of dimension $d=3$ and flavor $s=-1$ (panel (\textbf{a})), $s=0$ (panel (\textbf{b})), and $s=1$ (panel (\textbf{c}))  for  $n=0$ (blue solid lines), $n=1$ (red dashed lines), $n=2$ (yellow dotted lines) and $n=3$ (purple dot-dashed lines). Considered NGFs sizes are $N_{[0]}=2000$ nodes, $N_{[1]}=5994$ links, $N_{[2]}=5992$ triangles, and $N_{[3]}=1997$ tetrahedra. Adapted figure from \cite{torres2020simplicial}.
}
\label{Higher_spectral_dimension}
\end{figure} 
 In particular, the higher-order up-Laplacians of NGFs display a finite spectral dimension $d_S^{[n]}$ depending on the {order $n$}, the dimension of the simplicial complex $d$ and the flavor parameter $s$  \cite{torres2020simplicial}.  Therefore, we can define different spectral dimensions  for $0<n<d-1$. 
In order to show this remarkable geometrical property of NGFs in Figure $\ref{Higher_spectral_dimension}$  we provide numerical evidence of  the scaling  of   the cumulative density of non-zero eigenvalues $\rho_c^{\text{up}}(\lambda)$ of the ${\bf L}^{[\text{up}]}_{[n]}$  with  $\lambda$ for $\lambda\ll 1$, given by 
\bea
\rho_c^{[\text{up}]}(\lambda)\propto \lambda^{d_S^{[n]}/2},
\eea
 for different value of the   order $n$ of the up-Laplacian, and the flavor $s$ of the NGF.
 It follows that an NGF model is not characterized by a single spectral dimension, i.e.,  the spectral dimension $d_S^{[0]}$ of the graph Laplacian, rather the NGF simplicial complexes have a higher-order network geometry encoded in a vector of spectral dimensions 
 \bea
 {\bf d}_{S}=\left(d_S^{[0]},d_S^{[1]},d_S^{[2]}\ldots, d_S^{[d-1]}\right).
 \eea
Consequently, the diffusion dynamics defined on simplices of different order $n$ of the same NGF simplicial complex can be significantly different \cite{torres2015brain}.
We finally note  that for deterministic Apollonian and pseudo-fractal simplicial complexes that constitute the deterministic counterpart of NGF simplicial complexes, the higher-order spectral dimension can be predicted analytically by the real-space renormalization group \cite{reitz2020higher} showing that  the higher-order spectral dimension of these structures depends on the order $n$ and remains finite as long as  $0\leq n< d-1$.


\section{Simplicial synchronization}

Synchronization is a fundamental dynamical state  observed in  a wide variety of complex systems  and capturing among other phenomena important aspects of brain dynamics and circadian rhythms. 
The Kuramoto model \cite{kuramoto1975, strogatz2000kuramoto, acebron2005kuramoto, arenas2008synchronization, boccaletti2018synchronization} is a stylized model that  explains how
 coupled oscillators, that in absence of interactions would have different intrinsic frequencies, can start to follow a  collective coherent motion when their coupling constant $\sigma$, measuring the strength of their interaction, is larger than a critical value $\sigma_c$  also called {\em synchronization threshold}. 
 
In order to model the coupling between the oscillators the  Kuramoto model considers a  network of $N_{[0]}$ nodes and associates a phase $\theta_i$ to each node  $i\in \{1,2,\ldots, N_{[0]}\}$ of the network.
Therefore in the Kuramoto model, the dynamical state of the network is determined by the vector $\bm\theta$ of phases associated with its nodes given by 
\bea
\bm{\theta}=(\theta_1,\theta_2,\ldots, \theta_{N_{[0]}})^{\top}.
\label{theta}
\eea
Each phase $\theta_i$ describes the dynamical state of an oscillator that in absence of interactions oscillates at an intrinsic frequency $\omega_i$ drawn independently  from a distribution $g(\omega)$. Common choices for $g(\omega)$ are the unimodal Gaussian or Lorentzian distributions.

The equations determining the dynamics of the phases associated with the nodes include an important contribution indicating the coupling among the phases of neighbour nodes. This coupling term has a strength  modulated by the coupling constant $\sigma$.
In , it is assumed that this contribution expresses the tendency of the phase of any given node to oscillate together with the phases of its neighbour nodes.
The resulting standard Kuramoto dynamics is captured by the differential equations
 \bea
\dot {\theta}_i=\omega_i+\sigma \sum_{j=1}^{N_{[0]}} a_{ij}\sin \left(\theta_j-\theta_i\right),
\label{KG}
\eea
valid for every node $i$ of the network, where $a_{ij}$ is the generic element of the adjacency matrix of the network.
 The level of synchronization in the system is measured by the Kuramoto order parameter, 
\begin{equation}\label{eq:kuramoto_param_1d}
    Z_0= R_0\textrm{e}^{\mathrm{i}\Theta} = \frac{1}{N}\sum_{j=1}^{N_{[0]}} \textrm{e}^{\mathrm{i}\theta_j},
\end{equation}
where $R_0$ and $\Theta$ are both real  and where $0\leq R_0\leq 1$ measures the overall coherence and $\Theta=\Theta(t)$ is the phase of global oscillations.

The relation between the Kuramoto model and the graph Laplacian of the underlying network is revealed when the Kuramoto model is linearized for $|\theta_i-\theta_j|\ll 1$ for every pair of neighbour nodes $(i,j)$. In this limit the Kuramoto model can be shown to be described by the system of equations 
\bea
\dot {\bm\theta}=\bm\omega-\sigma {\bf L}_{[0]}\bm\theta,
\label{lin}
\eea
where $\bm{\omega}$ indicates  the vector of elements $\omega_i$ with $i\in \{1,2,\ldots, N_{[0]}\}$.

The Kuramoto model has been analytically solved only on a fully connected network, although important progress has been made  in understanding the Kuramoto model in random complex networks \cite{arenas2008synchronization,restrepo2005onset,chavez2005synchronization},  

In the fully connected network and in random networks  the Kuramoto model displays a second order phase transition at the synchronization threshold $\sigma=\sigma_c$
  when the number of nodes goes to infinity, i.e., $N_{[0]}\to \infty$. 
For  $\sigma<\sigma_c$  the Kuramoto model is in  an incoherent  state characterized by having a zero order parameter $R_0=0$. For $\sigma>\sigma_c$ the Kuramoto model is in a   coherent state characterized by a non-zero order parameter $R_0>0$  \cite{acebron2005kuramoto,strogatz2000kuramoto,arenas2008synchronization,boccaletti2018synchronization}.

In this chapter we show how the network geometry and topology of simplicial complexes, directly acting on the spectral properties of the graph Laplacian and the higher-order Laplacians,  can dramatically change the dynamical properties of the synchronization process on higher-order networks.
\begin{figure}[b]
\begin{center}
\includegraphics[scale=.25]{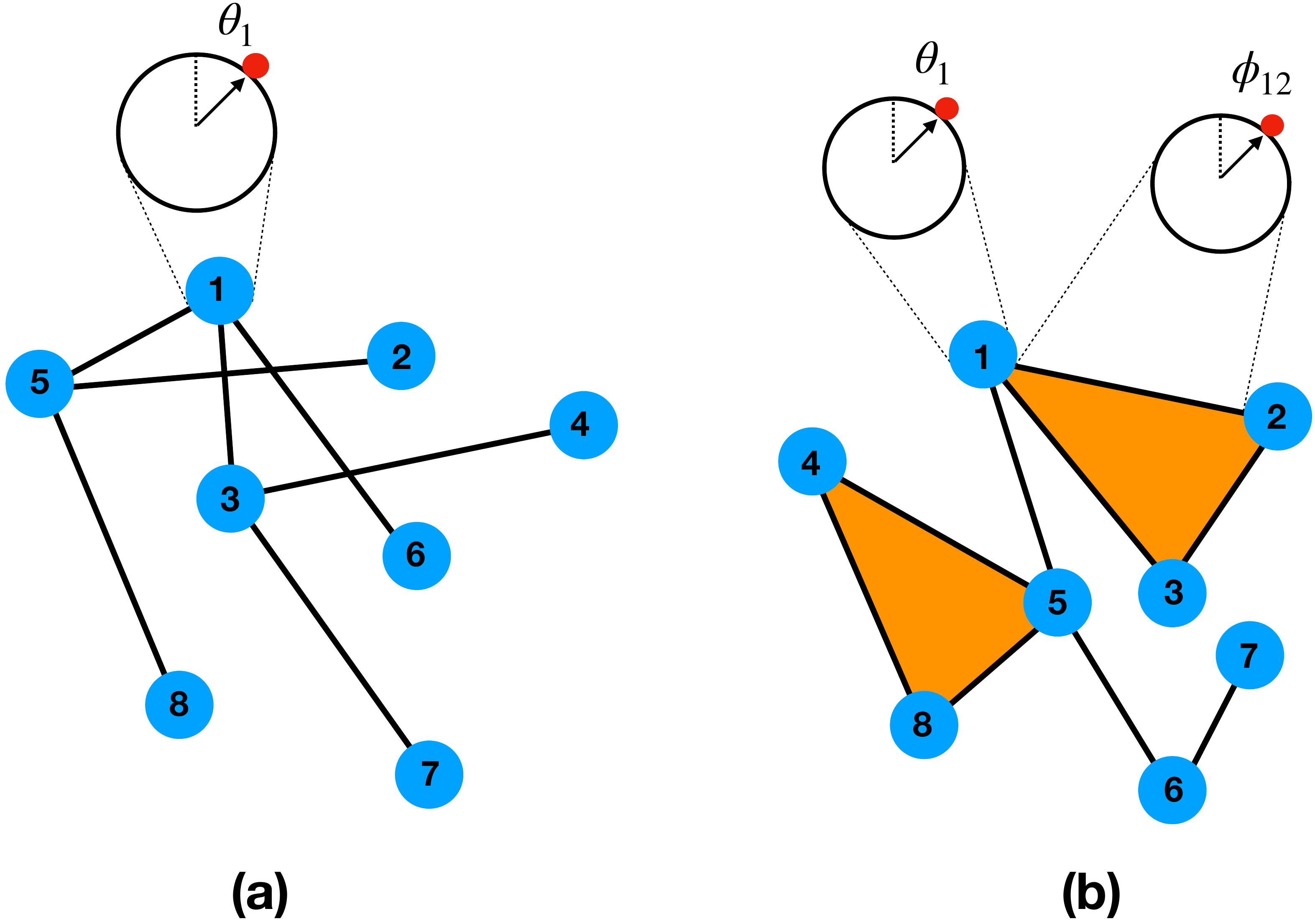}
\caption{ {\bf Schematic representation of the  Kuramoto model and the higher-order Kuramoto model capturing dynamics of topological signals.} The Kuramoto model (panel (\textbf{a})) captures the emergence of a synchronized state among coupled oscillators described by phases associated with the nodes of a network. The higher-order Kuramoto model  (panel (\textbf{b})) reveals the synchronization of topological signals on simplicial complexes, i.e., oscillators associated not only to the nodes of a simplicial complex, but also to higher-dimensional simplices such as links or triangles. Interestingly, topological signals of different dimension can co-exist and co-evolve and can be  non-trivially coupled leading to simultaneous explosive transitions. }
\end{center}
\label{fig:topo}       
\end{figure}
\begin{itemize}
\item{\bf Simplicial network geometry and the Kuramoto model.}
First we will illustrate how the phase diagram of the Kuramoto model changes when the model  is defined on the skeleton of simplicial complexes with distinct higher-order network geometry.
We have previously introduced  the finite spectral dimension $d_S^{[0]}$ of the graph Laplacian  as a  fundamental observable of the higher-order network geometry of networks and simplicial complexes. In the following we will discuss how the dynamics of the Kuramoto model depends on $d_S^{[0]}$ revealing that network geometries have a more rich dynamics than random networks.

\item{\bf Simplicial network topology  and the higher-order Kuramoto model.}
Secondly we will  reveal how the Kuramoto model can be considered as a limiting case of a much wider class of higher-order Kuramoto models describing coupling of topological signals. Topological signals are phases of oscillators associated  not only  to the nodes  but also to higher-order simplices of a simplicial complex.
For instance topological signals can be associated with both nodes and links of a simplicial complex as schematically described by Figure $\ref{fig:topo}$.
In this case the dynamical state of a simplicial complex is captured by the vector $\bm\theta$ of the phases associated with the nodes (defined in Eq. (\ref{theta})) and by the vector $\bm \phi$ of the phases associated with the links of the simplicial complex given by 
\bea
\bm{\phi}=\left(\phi_{\ell_1},\phi_{\ell_2},\ldots, \phi_{\ell_{N_{[1]}}}\right)^{\top}.
\eea
The phases associated with the links of the simplicial complexes are topological signals that have the potential to capture the dynamics of  fluxes in brain networks \cite{deville_brain} and biological transportation networks \cite{katifori1}. 
The newly formulated higher-order Kuramoto model opens new scenarios for characterizing how topology affects dynamics on higher-order networks and simplicial complexes.
As we will discuss in Sec. V the higher-order Kuramoto model has a linearized dynamics described by the higher-order Laplacians of the simplicial complex, and can display the simultaneous explosive synchronization transition of the soleinodal and the irrotational component of topological signals and even the simultaneous explosive synchronization transition of topological signals of different dimensions.
\end{itemize}

 \section{Kuramoto model on simplicial network geometry}

\subsection{Synchronization on simplicial network skeletons with finite spectral dimension}

In order to investigate the role that simplicial network geometry has on the Kuramoto model we explore the phase diagram  of the normalized Kuramoto model on networks with finite spectral dimension \cite{millan2019synchronization}.
Our theoretical results are then validated by simulations performed over the skeleton of the NGF with flavor $s=-1$. Indeed NGF with flavor $s=-1$  provide a very suitable benchmark to test our theoretical  results as they display a spectral dimension $d_S^{[0]}$ that can be changed by  tuning the dimension $d$ of the simplicial complex \cite{millan2018complex}.


The normalized Kuramoto model determines the dynamics of the phases $\bm\theta$ associated with the nodes of a network. The only difference with the standard Kuramoto model is that the coupling between the phase of a given node $i$ and the phases of its neighbour nodes is normalized with the node degree $k_i$.
Therefore, the normalized Kuramoto model is dictated by the differential equations
 \bea
\dot{\theta}_i=\omega_i+\sigma\sum_{j=1}^{N_{[0]}}  \frac{a_{ij}}{k_i}\sin \left(\theta_j-\theta_i\right),
\label{normalized}
\eea
where here and in the following we consider internal frequencies $\omega_i$ drawn independently from a normal distribution, i.e., $\omega_i\sim \mathcal{N}(0,1).$
The normalization of the coupling term by the degree of the node $i$ is  a very efficient way to screen out the effects of  the heterogeneity of the degrees of the nodes and single out only the effects due to the geometrical nature of the network of their interactions.

The linearized equation of the normalized Kuramoto model is therefore determined by the normalized Laplacian ${\hat{\bf L}}_{[0]}$ instead of the graph Laplacian ${\bf L}_{[0]}$, i.e.,
\bea
\dot {\bm\theta}=\bm\omega-\sigma {\bf \hat{L}}_{[0]} \bm\theta.
\label{lin_norm}
\eea

The analytical investigation  of the stability of the synchronized phase indicates that the spectral dimension $d_S^{[0]}$ of the (normalized) graph Laplacian of the network plays a fundamental role for determining the phase diagram of the normalized Kuramoto model in the limit of infinite network size $N_{[0]}\to \infty$ \cite{millan2019synchronization}.
Interestingly, we note that the spectral dimension of the graph Laplacian and the normalized graph Laplacian take the same value under very general regularity conditions of the network \cite{burioni1996universal}.
For the very heterogeneous NGF, numerical results show that the two spectral dimensions differ by a small amount as long as the topological dimension $d$ is small.
Depending on the value of the spectral dimension the phase diagram of the normalized Kuramoto model defined in the infinite network limit  changes drastically \cite{millan2019synchronization}:
\begin{itemize}
\item[(1)]
 For  networks  with finite spectral dimension  $d_S^{[0]}\leq2,$ the Kuramoto model  cannot synchronize and is found in the incoherent state for every value of the coupling constant $\sigma$.
\item[(2)] For networks with  spectral dimension $2<d_S^{[0]}\leq 4,$ global synchronization is not achievable in the infinite network limit but an entrained state can be observed.  Therefore it is possible the Kuramoto model can have a transition between an incoherent state and an entrained phase.
\item[(3)]
Only for networks with spectral dimension $d_S^{[0]}>4$ it is possible to see a synchronized phase.
\end{itemize}

These results reveal how the dynamics of the Kuramoto model  depends on the simplicial network geometry on which is defined,
and extend previous results valid on regular lattices of dimension $d$ \cite{hong2005collective, hong2007entrainment}.

\subsection{Frustrated synchronization on  Network Geometry with Flavor}

The NGF constitutes a perfect model to investigate numerically the role that simplicial network geometry has on the dynamics of the Kuramoto model \cite{millan2018complex,millan2019synchronization}.
Indeed, as we discussed previously, the NGF displays an emergent hyperbolic network geometry and for flavor $s=-1$ generates random hyperbolic manifolds.
The simplicial network geometry of NGF is also reflected on their spectral properties. Specifically, NGFs have a finite spectral dimension $d_S^{[0]}$ which for flavor $s=-1$ and $d\in \{2,3,4\}$ can be approximated by $d_S^{[0]}\simeq d$. It follows that by changing the dimension $d$ of the NGF with flavor $s=-1$ we can explore the dynamics of the Kuramoto model when the global synchronization state is not stable in the infinite network limit.

A computational finite size analysis of the Kuramoto model \cite{millan2018complex,millan2019synchronization} reveals, in agreement with the theoretical expectations, that for $d=2$ and $d_S^{[0]}\simeq 2$, global synchronization is never achieved for large network sizes. 
On the other hand, synchronization in NGFs with $d=3$ and $d=4$  and $d_S^{[0]}\simeq d$ is only possible for finite networks. In fact  its onset occurs  for higher couplings when the system size  increases, revealing, in agreement with the theoretical expectation, that in the limit $N_{[0]}\to \infty$ this state is never achieved.

Interestingly, we observe that for NGF with $d=3$ and $d=4$  and $d_S^{[0]}\simeq d$ the Kuramoto model exhibits a phase with entrained synchronization that we call a phase of {\em frustrated synchronization} for a wide range of coupling values. In the {\em frustrated synchronization regime} the order parameter $R_0$ displays strong temporal fluctuations. This phase is observed on finite NGF between the incoherent state and the globally synchronized state,  and for $d=3$ a much broader regime of large fluctuations is observed  than for the $d=4$ case.

Interestingly, the frustrated synchronization phase of NGFs is not only  characterized by strong temporal fluctuations of the global order parameter but displays also strong spatial fluctuations induced by the non-trivial hyperbolic network geometry of the NGF. As such, the frustrated synchronization phase can capture an important mechanism for inducing spatio-temporal fluctuations in brain dynamics. 

\begin{figure}
    \centering
    \includegraphics[width = 1.0\textwidth]{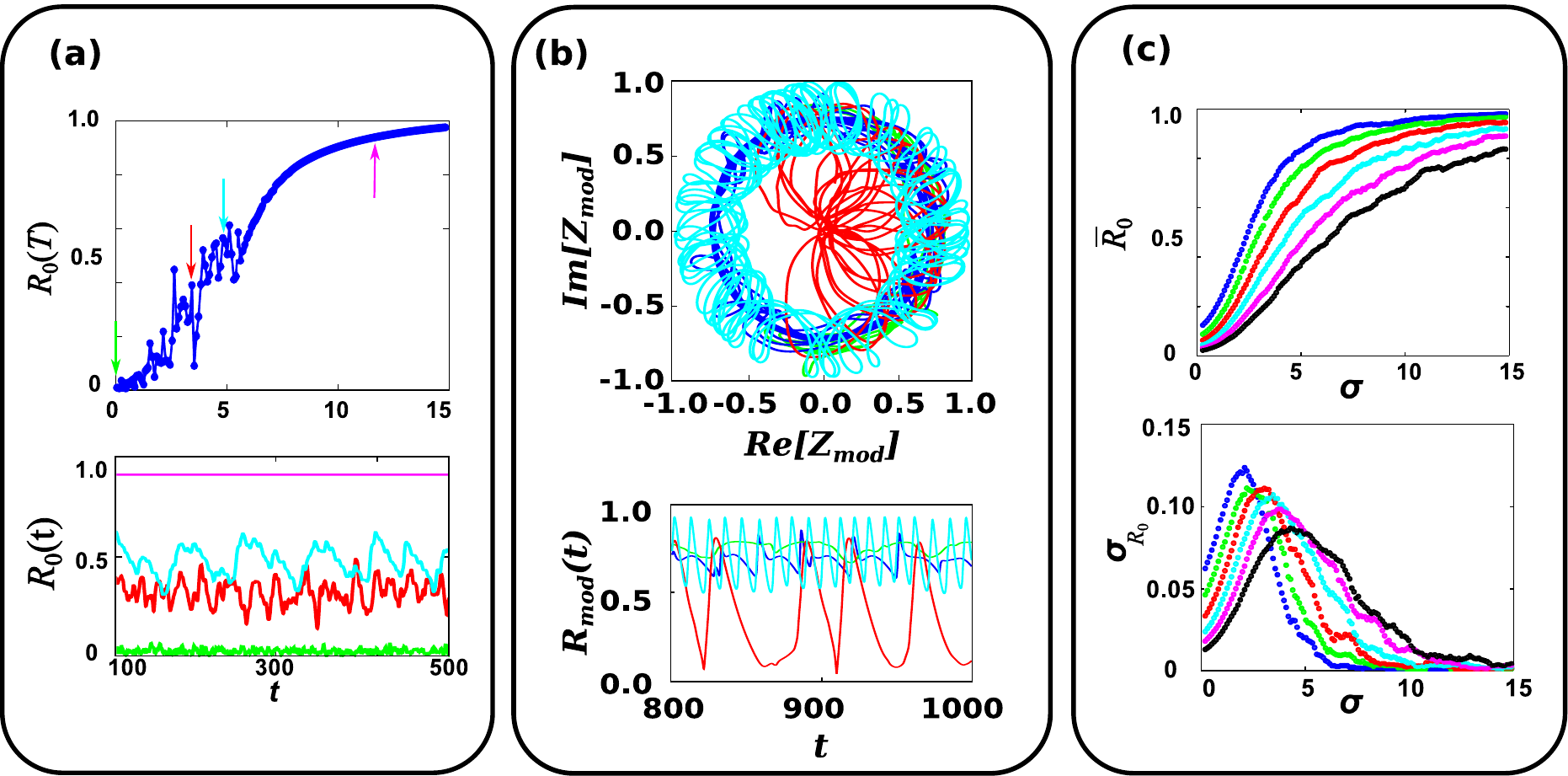}
    \caption{\textbf{Frustrated synchronization on NGF characterized by spatio-temporal fluctuations of the order parameter.} 
 {\bf Panel (a)} The synchronization order parameter $R_0(T)$ calculated at time $T$  is plotted versus the coupling constant  $\sigma$ revealing the regime of frustrated synchronization (top panel). 
The time series $R_0(t)$ are shown (bottom panel) for the values of the   coupling constant $\sigma$ indicated by arrows in the top panel. These time series reveal the temporal fluctuations of the order parameter in the frustrated synchronization regime.
    \textbf{Panel (b)} The spatial fluctuations of the order parameter in the frustrated synchronization regime are revealed by  the local order parameter $Z_{\textrm{mod}}$ of four different communities of the NGF calculated for $\sigma=5$ (top panel).  In this representation, a circular trajectory describes a situation of global oscillations of the nodes in the community, with constant $R_\textrm{mod}(t)$.   Random trajectories around the origin describe unsynchronized communities. Partially synchronized communities, on the other hand, may describe more complex trajectories. The bottom panel shows the corresponding time series of $R_{\textrm{mod}}(t)$. 
    \textbf{Panel (c)} Synchronization transition as given by the mean order parameter $\bar{R}_0$ averaged over time and its variance $\sigma_{R_0}$, as functions of $\sigma$, for different network sizes $N=100$ (blue), $200$, $400$, ..., $3200$ (black).
    The apparent onset of the synchronized regime is retarded to large values of the coupling constant $\sigma$ for large network sizes, revealing that the NGF of dimension $d=3$ cannot sustain a synchronized phase in the limit of infinite network sizes.
    All simulations reported in the figures are obtained for NGF of flavor $s=-1$, dimension $d=3$ and number of nodes $N_{[0]}=1600$. In panel (\textbf{a}) $T=500$. Figure adapted from \cite{millan2018complex}. }
    \label{fig:s4_fig1}
\end{figure}

The hyperbolic network geometry of  NGF  has a strong hierarchical nature that is responsible for the emergence of a relevant  community structure. 
In order to study how the dynamics of the Kuramoto model is affected by the community structure of the NGFs, 
we define  mesoscopic synchronization order parameters $Z_{\textrm{mod}}$ that characterize the  dynamical state of each community: 
\begin{equation}
    Z_\textrm{mod}  = R_\textrm{mod} \textrm{e}^{\mathrm{i}\Theta_\textrm{mod}} = \frac{1}{\left|\mathcal{C}\right|} \sum_{j\in \mathcal{C}} \textrm{e}^{\mathrm{i}\theta_j},
\end{equation}
where $\mathcal{C}$ is the set of nodes in the  community and $\left|\mathcal{C}\right|$ the total number of nodes in said community. 
Figure $\ref{fig:s4_fig1}$b displays the trajectory of $Z_\textrm{mod}=Z_\textrm{mod}(t)$ in the complex plane for some exemplary modules of an NGF with flavour $s=-1$ in $d=3$, for the coupling that leads to the largest fluctuations of $R_0$ as a function of time.
As shown in the figure, different modules display different synchronization regimes and may oscillate at different frequencies. Due to the underlying geometrical structure of NGFs, these modules correspond to spatially localized regions.

We note here that the frustrated synchronization observed in NGF can be related with analogous phases observed in other hierarchical models 
\cite{moretti2013griffiths,villegas2014frustrated} where temporal fluctuations of the synchronization order parameter are observed. However, the combination of both temporal and spatial fluctuations is a specific property of the frustrated synchronization in NGF due to their rich simplicial network geometry.

\section{Higher-order Kuramoto model: a topological approach to synchronization}

\subsection{Synchronization of topological signals}

Simplicial complexes are formed by nodes and higher-order simplices including links, triangles, tetrahedra, and so on. As such, simplicial complexes have the ability to sustain topological signals, i.e., dynamic variables not only associated with the nodes of their network skeleton but also associated with links, triangles, and so on \cite{millan2020explosive,ghorbanchian2020higher,Barbarossa}.
Topological signals have the ability to capture dynamics associated with links, such as fluxes in brain networks \cite{deville_brain} and biological transportation networks \cite{katifori1,katifori2}.

Here we will present the  higher-order Kuramoto model  \cite{millan2020explosive} that reveals how topological signals can undergo continuous and discontinuous synchronization transitions. Interestingly, we will observe  that this synchronization transition can be detected only if the signals are filtered with the  appropriate topological operators. Therefore, the model not only captures a new topological critical phenomenon but also prescribes a way to process real data in order to investigate whether this topological synchronization phenomenon can be observed in real systems such as the brain or biological transportation networks.

\subsection{Higher-order Kuramoto dynamics}

The higher-order Kuramoto model \cite{millan2020complex}  describes the synchronization of topological signals defined on simplices of dimension $n$. For instance, one can consider topological signals defined on the links of a simplicial complex (case $n=1$) or alternatively one can consider signals defined on the triangles of a simplicial complex (case $n=2$). The higher-order Kuramoto model is the most natural extension of the Kuramoto model to capture  the synchronization of higher-order topological signals. 

The standard  Kuramoto dynamics describes the dynamics of the phases $\bm\theta$ associated with the nodes of the network. This dynamics, defined by Eq. (\ref{KG}), can be expressed in terms of the incidence matrix $\bm{B}_{[1]}$ (see Appendix) as 
\bea
\dot{\bm{\theta}} = \bm{\omega} - \sigma \bm{B}_{[1]} \sin \bm{B}_{[1]}^T \bm{\theta}.
\label{Kuramoto_B}
\eea

The higher-order Kuramoto dynamics  describes instead the dynamics of topological signals $\bm\phi$ with $\phi_\alpha$ indicating the phase associated with the simplex $\alpha$ of dimension $n>0$.
Therefore, using the insights coming from algebraic topology, the natural definition of the {\em simple higher-order Kuramoto model} is 
\bea\label{eq:kuramo_HO}
\dot{\bm{\phi}} = \bm{\hat\omega} - \sigma \bm{B}_{[n+1]} \sin \bm{B}_{[n+1]}^T \bm{\phi}  - \sigma \bm{B}_{[n]}^T \sin \bm{B}_{[n]} \bm{\phi}, 
\eea
where $\bm{\hat\omega}$ is the vector of intrinsic frequencies $\hat{\omega}_\alpha$ associated with each $n$-dimensional simplex $\alpha$ drawn independently from a normal distribution, i.e., $\hat{\omega}_\alpha\sim \mathcal{N}(\Omega_1,1/\tau_1)$.
As the standard Kuramoto model can be related to the graph Laplacian via linearization (see Eq. (\ref{lin})), the higher-order Kuramoto model can be related to the higher-order Laplacian upon linearization, leading to 
\bea
\dot{\bm{\phi}} = \bm{\hat\omega} - \sigma {\bf L}_{[n]}\bm \phi.
\eea
The higher-order Kuramoto dynamics defined on topological signals $\bm\phi$ associated with $n$ dimensional simplices can be projected on $(n+1)$ and $(n-1)$ dimensional simplices by applying to the signals the incidence matrices. For instance a dynamics defined on topological signals  associated with links can be projected on  nodes or on triangles.  
Specifically, we have that the projected dynamics $\bm\phi^{[-]}$ on $(n-1)$-dimensional simplices  and the projected dynamics $\bm\phi^{[+]}$ on $(n+1)$-dimensional simplices is given by 
\bea
\bm\phi^{[-]}&=&{\bf B}_{[n]}\bm\phi,\nonumber \\
\bm\phi^{[+]}&=&{\bf B}_{[n+1]}^{\top}\bm\phi,
\eea
where, for $n=1$, ${\bf B}_{[n]}$ indicates the discrete divergence and ${\bf B}_{[n+1]}^{\top}$ indicates the discrete curl. Therefore $\bm{\phi}^{[-]}$ indicates the irrotational component of $\bm\phi$ while $\bm\phi^{[+]}$ indicates the solenoidal component of $\bm\phi$.
For the simple higher-order Kuramoto model defined in Eq. (\ref{eq:kuramo_HO}), by recalling that ${\bf B}_{[n+1]}^{\top}{\bf B}_{[n]}^{\top}={\bf 0}$ and that  ${\bf B}_{[n]}{\bf B}_{[n+1]}={\bf 0},$ it follows that the projected topological signals $\bm{\phi}^{[-]}$ and $\bm{\phi}^{[+]}$ obey the uncoupled system of equations
\bea 
    \dot{\bm{\phi}}^{[-]} &=& \bm{B}_{[n]} \bm{\hat\omega} - \sigma  \bm{L}_{[n-1]}^{[\textrm{up}]} \sin \left( \bm{\phi}^{[-]} \right), \nonumber\\
    \dot{\bm{\phi}}^{[+]} &=& \bm{B}^T_{[n+1]} \bm{\hat\omega} - \sigma \bm{L}_{[n+1]}^{[\textrm{down}]} \sin \left( \bm{\phi}^{[+]} \right).
 \label{eq:dynamics_proj_simple}
\eea
Therefore, the solenoidal and the irrotational components of the topological signals are decoupled for the simple higher-order Kuramoto model.

\begin{figure}
    \centering
    \includegraphics[width = 1.0\textwidth]{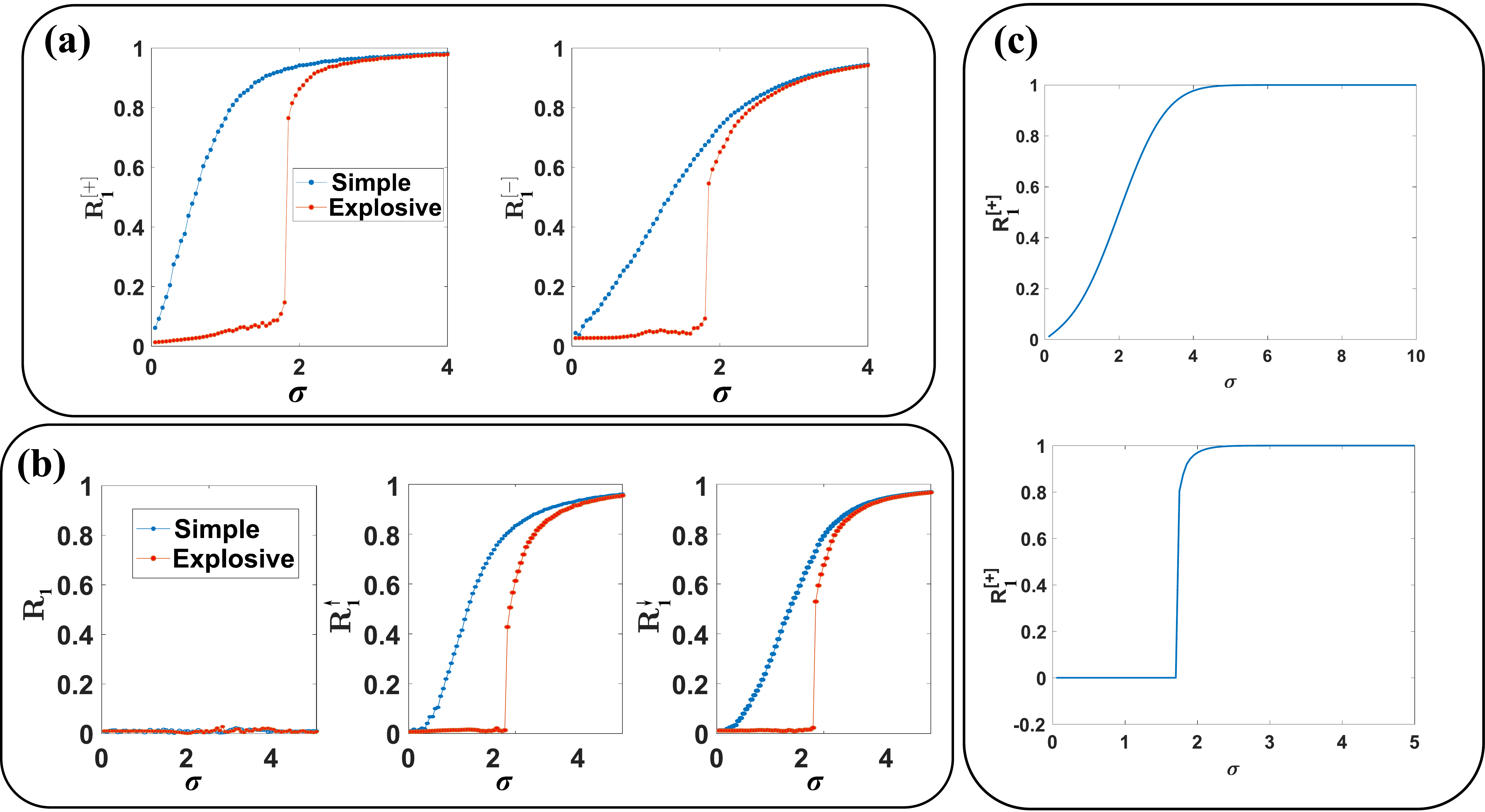}
    \caption{ \textbf{Higher-order Kuramoto dynamics.}
        The synchronization of higher-order topological signals is captured by the simple and explosive higher-order Kuramoto models.
        \textbf{Panel (a)} The order parameters $R_1^{[+]}$ and $R_1^{[-]}$ reveal the synchronization transition of topological signals defined on links ($n=1$) of the configuration model of simplicial complexes. 
   \textbf{Panel (b)} The na\"ive order parameter  $R_1$ does not reveal the synchronization  transition of topological signals defined on the links of the configuration model of simplicial complex, while the order parameters $R_{1}^{\uparrow}$ and $R_1^{\downarrow}$, sensible to the irrotational and solenoidal  decomposition of the signal, do reveal the transition (as well as $R_1^{[+]}$ and $R_1^{[-]}$ shown  in panel (\textbf{a})).
    The underlying network is the same for both panels and has $N_{[0]}=1000$ nodes, $N_{[1]}=5299$ links and $N_{[2]} = 4147$ triangles. The generalized degree of the nodes is power-law distributed with power-law exponent $\gamma=2.8$.
  \textbf{Panel (c)} shows  the theoretical expectations provided by an effective phenomenological model treated in Ref. \cite{millan2020explosive} for  the simple (top) and explosive (bottom) higher-order Kuramoto models. 
Reprinted figure with permission from Ref. \cite{millan2020explosive}. \copyright   Copyright  (2020)  by  the American Physical Society.}
    \label{fig:s5_fig1}
\end{figure}

The higher-order Kuramoto dynamics is remarkable from two perspectives:
\begin{itemize}
\item[(1)] First of all it uses topology to naturally define the higher-order interactions between the topological signals.
Indeed the incidence matrices define higher-order interactions with a clear prescription indicating the coupled variables and the sign of their interactions.
For example, the  higher-order Kuramoto dynamics for  $n=1$ dimensional simplices of the  simplicial complex shown in Figure $\ref{fig:4}$  reads 
\bea
\dot {\theta}_{[1,2]}&=&\hat\omega_{[1,2]}-\sigma \sin(\theta_{[2,3]}-\theta_{[1,3]}+\theta_{[1,2]})-\sigma\left[\sin(\theta_{[1,2]}-\theta_{[2,3]})+\sin(\theta_{[1,3]}+\theta_{[1,2]})\right],\nonumber\\
\dot {\theta}_{[1,3]}&=&\hat\omega_{[1,3]}+\sigma \sin(\theta_{[2,3]}-\theta_{[1,3]}+\theta_{[1,2]})-\sigma\left[\sin(\theta_{[1,3]}+\theta_{[1,2]})+\sin(\hat{\theta}_{[3]})\right],\nonumber\\
\dot {\theta}_{[2,3]}&=&\hat\omega_{[2,3]}-\sigma \sin(\theta_{[2,3]}-\theta_{[1,3]}+\theta_{[1,2]})-\sigma \left[\sin(\theta_{[2,3]}-\theta_{[1,2]})+\sin(\hat{\theta}_{[3]})\right],\nonumber \\
\dot {\theta}_{[3,4]}&=&\hat\omega_{[3,4]}-\sigma \left[\sin(\theta_{[3,4]})-\sin(\hat{\theta}_{[3]})\right],
\eea
with $\hat{\theta}_{[3]}$ indicating the three-body interaction
\bea
\hat{\theta}_{[3]}=\theta_{[13]}+\theta_{[23]}-\theta_{[34]}.
\eea
Therefore, the choice of the higher-order interactions in the higher-order Kuramoto model is naturally dictated by topology.
\item[(2)]
The synchronization of the higher-order Kuramoto model is only detectable if the right topological filtering of the data is performed.
Indeed the  na\"ive order parameter  
\bea
R_n = \frac{1}{N_{[n]}} \left| \sum_{\alpha=1}^{N_{[n]}} e ^ {\mathrm{i}\phi_\alpha}  \right|\eea associated with the unfiltered topological signal $\bm\phi$ does not detect any synchronization transition (see Figure $\ref{fig:s5_fig1}$).
Instead, the order parameter associated with the solenoidal and the irrotational components of the topological signal do detect the synchronization transition of the topological signals (see Figure $\ref{fig:s5_fig1}$). These order parameters are given by 
\bea
R_n^{[-]} = \frac{1}{N_{[n-1]}} \left| \sum_{\alpha=1}^{N_{[n-1]}} e ^ {\mathrm{i}\phi_\alpha^{[-]}}  \right|,\ \ \ 
R_n^{[+]} = \frac{1}{N_{[n+1]}} \left| \sum_{\alpha=1}^{N_{[n+1]}} e ^ {\mathrm{i}\phi_\alpha^{[+]}}  \right|,
\eea
or alternatively by 
\bea
R_n^{\downarrow} = \frac{1}{N_{[n]}} \left| \sum_{\alpha=1}^{N_{[n]}} e ^ {\mathrm{i}\phi_\alpha^{\downarrow}}  \right|,\ \ \ 
R_n^{\uparrow} = \frac{1}{N_{[n]}} \left| \sum_{\alpha=1}^{N_{[n]}} e ^ {\mathrm{i}\phi_\alpha^{\uparrow}}  \right|,
\eea
where $\bm\phi^{\downarrow}={\bf L}_{[n]}^{[\text{down}]}\bm\phi$ and $\bm\phi^{\uparrow}={\bf L}_{[n]}^{[\text{up}]}\bm\phi$. 
\end{itemize}

The synchronization transition described by the simple higher-order Kuramoto model leads to a continuous transition occurring at zero coupling, i.e., the synchronization threshold is $\sigma_c=0$ as long as $n>0$. 
However, the higher-order Kuramoto model admits a formulation called {\em explosive higher order Kuramoto model} that displays instead a discontinuous transition at a non zero coupling $\sigma_c>0$.

The {\em explosive  higher-order Kuramoto dynamics} \cite{millan2020explosive} implements an adaptive coupling of the projected dynamics  of $\bm{\phi}^{[+]}$ and $\bm{\phi}^{[-]}$ through their global order parameters.  The adopted  adaptive coupling is inspired by analogous couplings previously applied to multilayer and simple networks \cite{zhang2015explosive}. The explosive higher-order Kuramoto model \cite{millan2020explosive} is defined by the system of equations
\begin{equation}\label{eq:dynamics_exp}
    \dot{\bm{\phi}} = \bm{\hat\omega} -\sigma R^{[-]}\bm{B}_{[n+1]} \sin \bm{B}^T_{[n+1]} \phi -\sigma R^{[+]} \bm{B}^T_{[n]} \sin \bm{B}_{[n]} \phi.
\end{equation}
It follows that the dynamics projected on the $(n+1)$ and $(n-1)$-dimensional simplices now obeys the coupled system of equations
\bea 
    \dot{\bm{\phi}}^{[+]} &=& \bm{B}^T_{[n+1]} \bm{\hat\omega} - \sigma R^{[-]} \bm{L}_{[n+1]}^{[\textrm{down}]} \sin \left( \bm{\phi}^{[+]} \right), \nonumber \\
    \dot{\bm{\phi}}^{[-]} &=& \bm{B}_{[n]} \bm{\hat\omega} - \sigma R^{[+]} \bm{L}_{[n-1]}^{[\textrm{up}]} \sin \left( \bm{\phi}^{[-]} \right). \label{eq:dynamics_proj_expl2}
\eea

Numerical simulations on the configuration model of simplicial complexes  with power-law distribution of generalized degrees reveal that the explosive higher-order Kuramoto model displays a discontinuous phase transition. The nature of the transition confirms the theoretical expectations obtained with an approximate phenomenological approach.
This transition is clearly detected by a discontinuity  in $R^{[+]}_n$ and $R^{[-]}_n$  and in  $R^{\downarrow}_n$ and $R^{\uparrow}_n$ as well, but is not captured by the na\"ive order parameter $R_n$ (see Fig. $\ref{fig:s5_fig1}$). 

In  Ref. \cite{millan2020explosive} it has been shown that  the nature of the phase transition does not change if the generalized degree distribution is more uniform or if the simplicial complex has a non trivial network geometry.
Interestingly, simple and explosive higher-order Kuramoto models can be investigated  on simplicial complexes constructed from real connectomes leading to continuous (for the simple higher-order Kuramoto model) and discontinuous (for the explosive higher-order Kuramoto model) synchronization transitions. 

\subsection{Coupled topological signals}

So far we have considered the synchronization of topological signals defined on simplices of dimension $n> 0$. However, topological signals associated with simplices of different dimension can co-exist and co-evolve. For instance, phases associated with the nodes of a simplicial complex can be coupled with phases associated with its links.
  In this section we will show how  different topological signals, i.e., phases defined on simplicial complexes of different dimensions, can be coupled to each other leading to simultaneous  explosive synchronization transitions. For simplicity of presentation we will focus on phases defined on nodes and links, but we emphasize that our formalism allows one to consider the interaction of more general topological signals.

\begin{figure}[t]
\sidecaption[t]
    \centering
    \includegraphics[width = 1.0\textwidth]{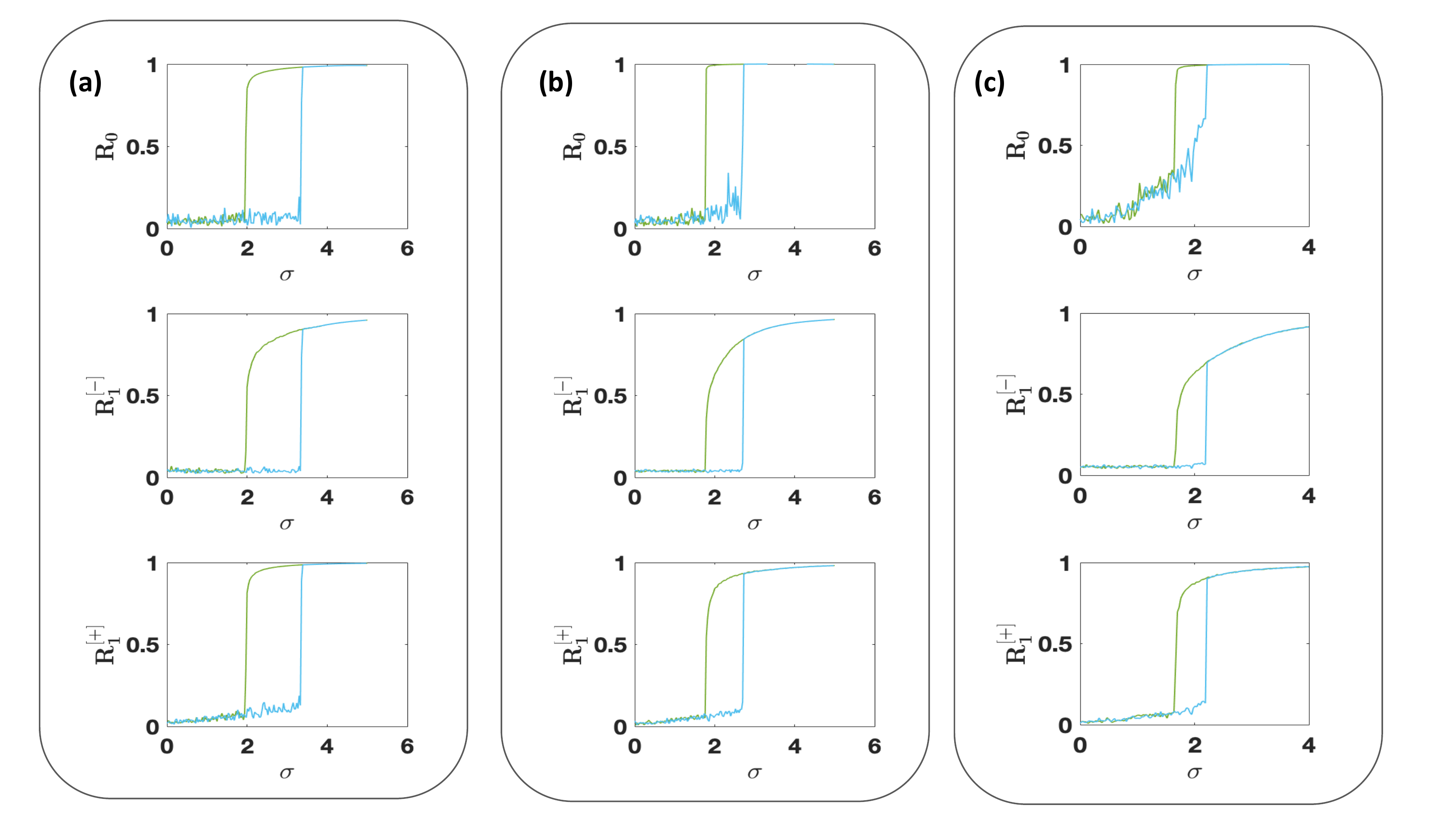}
\caption{ {\bf Simultaneous explosive synchronization of coupled topological signals.} Model 1 displays the simultaneous explosive (discontinuous) transition captured by the order parameters $R_0$, $R_1^{[-]}$ and $R^{[+]}_1$.
\textbf{Panels (a), (b), and (c)} provide numerical evidence of this discontinuous transition on different simplicial network topologies: the NGF with flavor $s=-1$ and dimension $d$ (panel (\textbf{a})), the configuration model of pure $d=3$ simplicial complex with power-law generalized degree distribution with power-law exponent $\gamma=2.8$ (panel (\textbf{b})) and the clique complex of the Caenorhabditis elegans (C.elegans) connectome coming from Ref. \cite{celegans} (panel (\textbf{c})).
Figure adapted from $\cite{ghorbanchian2020higher}$.}
\label{fig:coupled}       
\end{figure}

We start by considering Model 1, an explosive higher-order Kuramoto model of coupled signals of nodes and links. This model differs from the explosive higher-order Kuramoto model defined in the previous section as it includes an additional  adaptive coupling  among the topological signals of nodes and links, and obeys the equations
\bea
\dot{\bm{\theta}}&=& \bm{\omega}-\sigma R_1^{[-]}{\bf B}_{[1]}\sin ({\bf B}^{\top}_{[1]}\bm{\theta}),\nonumber \\
\dot{\bm{\phi}}&=& \hat{\bm{\omega}}-\sigma R_0R_1^{[+]} {\bf B}^{\top}_{[1]}\sin ({\bf B}_{[1]}\bm{\phi})-R_1^{[-]}\sigma {\bf B}_{[2]}\sin ({\bf B}_{[2]}^{\top}\bm{\phi}).
\label{model1}
\eea
This model simulated in a wide variety of simplicial complexes including the NGF, the configuration model of simplicial complexes, and the clique complex of real connectomes displays a simultaneous explosive (i.e., discontinuous) synchronization of the topological signals defined on nodes and of the soleinodal and irrotational component of the topological signals defined on links \cite{ghorbanchian2020higher}.
Indeed, at a critical threshold $\sigma=\sigma_c$ we observe a discontinuity in the three order parameters $R_0,R_1^{[-]}$ and $R_1^{[+]}$. In  Figure $\ref{fig:coupled}$ we present numerical evidence of this discontinuous transition by displaying the corresponding hysteresis loop in the order parameters. In particular, instead of plotting the order parameters obtained at each value of $\sigma$ starting from random initial conditions as in Figure $\ref{fig:s5_fig1}$ here we display the order parameters along  the forward and backward synchronization transitions obtained by first adiabatically  increasing and then decreasing the coupling constant $\sigma$. 

Interestingly, this model (Model 1) admits a variation (Model 2) that describes  the simultaneous synchronization of the topological signals associated with the nodes and the irrotational component of the topological signal associated with the links. This simpler version of the explosive higher-order Kuramoto model of coupled topological signals can be also defined on pairwise networks and is amenable to an exact analytical treatment on fully connected networks and to an accurate annealed approximation solution on random networks with a given degree distribution.
More specifically, Model 2 only couples the signal of the nodes with the signal of the links projected to the nodes, as described by the differential equations  
  \bea
\dot{\bm{\theta}}&=& \bm{\omega}-\sigma R_1^{[-]}{\bf B}_{[1]}\sin ({\bf B}^{\top}_{[1]}\bm{\theta}),\\
\dot{\bm{\phi}}&=& \hat{\bm{\omega}}-\sigma R_0 {\bf B}^{\top}_{[1]}\sin ({\bf B}_{[1]}\bm{\phi})-\sigma {\bf B}_{[2]}\sin ({\bf B}_{[2]}^{\top}\bm{\phi}).
\label{model2}
\eea

\begin{figure}[t]
\sidecaption[t]
    \centering
    \includegraphics[width = 1.0\textwidth]{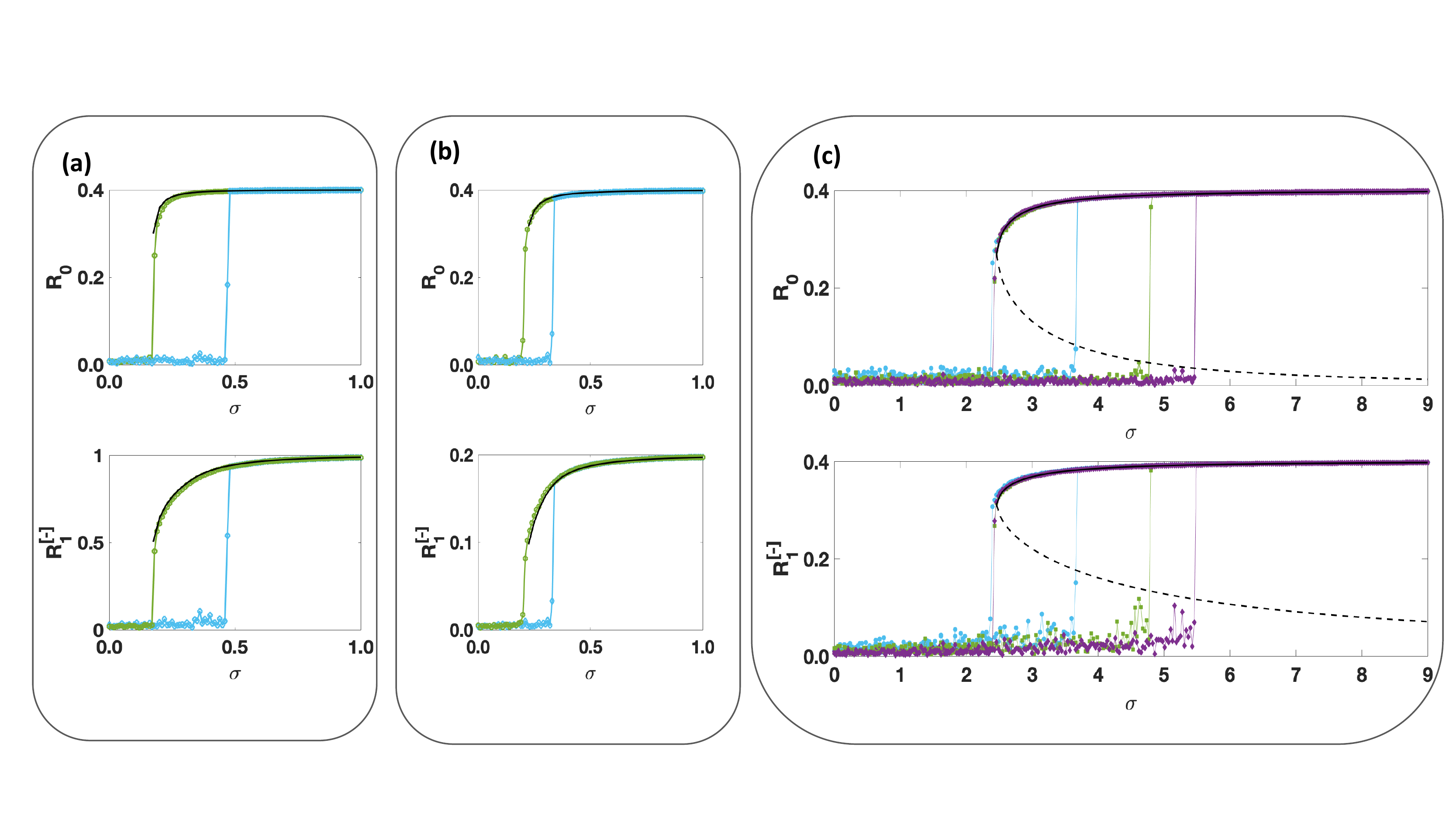}
\caption{{\bf Theoretical prediction for Model 2 of explosive higher-order Kuramoto model of coupled dynamical signals applied to random and fully connected networks.} The hysteresis loop for Model 2  is shown for different types of networks. The backward transition is fully captured by the theoretical expectations (solid black lines).
\textbf{Panels (a), (b)} and \textbf{(c)} show  the order parameters $R_1$ and $R_1^{[-]}$ versus the coupling constant $\sigma$ for: A Poisson network with average degree $c=12$ and $N_{[0]}=1600$ nodes (panel (\textbf{a})), a scale-free network with minimum degree $m=6$, power-law exponent $\gamma=2.5$ and $N_{[0]}=1600$ nodes (panel (\textbf{b})) and a fully connected network of different network sizes $N_{[0]}=500$ (cyan symbols), $N_{[0]}=1000$ (green symbols)and $N_{[0]}=2000$ (purple symbols) (panel (\textbf{c})). From panel (\textbf{c}) it is evident that the forward transition  occurs at higher values of $\sigma$ for larger network size, confirming the theoretical prediction  indicating that the transition is driven by finite size effects and  it is  absent in the infinite network limit. Figures adapted from $\cite{ghorbanchian2020higher}$.}
\label{fig:transitions}       
\end{figure}

Projecting the dynamics of the link phases down to 0-simplices as in the previous section, we introduce $\bm \psi$ for simplicity of notation with 
\bea
{\bm \psi} \equiv {\bm \phi}^{[-]} = {\bf B}_{[1]} {\bm \phi}.
\eea
 By left multiplying Eq.~(\ref{model2}) by ${\bf B}_{[1]}$, we obtain the closed system of equations for ${\bm \theta}$ and ${\bm \psi}$
\bea
\dot{\bm{\theta}}&=& \bm{\omega}-\sigma R_1^{[-]}{\bf B}_{[1]}\sin ({\bf B}^{\top}_{[1]}\bm{\theta}),\nonumber \\
\dot{\bm{\psi}}&=& \tilde{\bm{\omega}}-\sigma R_0 {\bf L}_{[0]}\sin ({\bm \psi}),
\label{model2projected}
\eea
where $\tilde{\bm{\omega}} = {\bf B}_{[1]} \hat{\bm \omega}$. Here we assume $\omega\sim \mathcal{N}(0,1)$ and $\hat{\omega}_{\alpha}\sim \mathcal{N}(\Omega_1,1/\tau_1)$. With these hypotheses  the internal frequencies of the links projected on the nodes $\{\tilde{\omega}_{i}\}_{i=1,2,\ldots, N_{[0]}}$  are  Gaussian correlated variables with average 
\bea
\Avg{\tilde{\omega}_i}=\left[\sum_{j<i}a_{ij}-\sum_{j>i}a_{ij}\right]\Omega_1
\label{avgomega}
\eea
and with correlation matrix ${\bf C}$ of elements $
C_{ij}=\Avg{\tilde{\omega}_i\tilde{\omega}_j}-\Avg{\tilde{\omega}_i}\Avg{\tilde{\omega}_j}$ given by 
\bea
{\bf C}={\bf L}_{[0]}\frac{1}{\tau_1^2}.
\eea
 To understand the nature of the synchronization transition analytically when Model 2 is defined  on an uncorrelated random graph, in the following we discuss the solution of the model in the  {\it annealed approximation}.
 The annealed approximation is a widely used approximation to study dynamical processes on random uncorrelated networks which consists in substituting the adjacency matrix entries of the  network $a_{ij}$ by their average values in an uncorrelated random  network with given degree sequence, i.e.,
 \bea
 a_{ij}\to \frac{k_i k_j}{\avg{k}N}.\eea 
 Using this approximation,  Eqs. (\ref{model2projected}) can be recast into the differential equations
\bea
\dot{\bm \theta}&=&\bm\omega-\sigma R_1^{[-]}\hat{R}_0{\bf k}\cdot \sin(\bm{\theta}-\hat\Theta),\nonumber  \\
\dot{\bm\psi}&=&\tilde{\bm\omega}+\sigma  R_0 \hat{R}_1^{[-]}{\bf k}\sin\hat\Psi-\sigma R_0 {\bf k}\odot \sin \bm\psi, \label{psip}
\eea
where $\odot$ indicates the Hadamard product (element by element multiplication) and 
where   two auxiliary  complex order parameters are defined as 
\bea
\hat{R}_0e^{\mathrm{i}\hat\Theta}=\sum_{i=1}^{N_{[0]}}\frac{k_i}{\avg{k}{N_{[0]}}}e^{\mathrm{i}\theta_i}, \quad
\hat{R}_1^{[-]}e^{\mathrm{i}\hat\Psi}=\sum_{i=1}^{N_{[0]}}\frac{k_i}{\avg{k}{N_{[0]}}}e^{\mathrm{i}\psi_i},
\eea 
with $\hat{R}_0,\hat\Theta,\hat{R}_1^{{[-]}}$ and $\hat\Psi$ being real.
Let us  indicate with $g(\omega)$ the probability distribution of the internal frequencies of the nodes and with $G_i(\tilde{\omega})$ the marginal probability distribution of the internal frequencies of the links projected on node $i$, i.e., the marginal probability that $\tilde{\omega}_i=\tilde{\omega}$. With this notation it is possible to derive the analytic solution of Eqs. (\ref{psip})  which gives the following expression for the order parameters
\bea
R_0=\frac{1}{{N_{[0]}}}\sum_{i=1}^{N_{[0]}} r_0(i), &\quad &\hat{R}_n=\sum_{i=1}^{N_{[0]}}\frac{k_i}{\avg{k}{N_{[0]}}} r_0(i),\nonumber \\
R_1^{[-]}=\frac{1}{{N_{[0]}}}\sum_{i=1}^{N_{[0]}} r_1^{[-]}(i), &\quad &\hat{R}_1^{[-]}=\sum_{i=1}^{N_{[0]}}\frac{k_i}{\avg{k}{N_{[0]}}} r_1^{[-]}(i),\nonumber \\
\eea
with ${r}_0(i)$ and ${r}_1^{[-]}(i)$ given by 
\bea
{r}_0(i)&=&\int_{|\hat{c}_i|<1} d\omega g(\omega) \sqrt{1-\left(\frac{\omega-\Omega_0}{\sigma k_i \hat{R}_0R_1^{[-]}}\right)^2},\nonumber \\
{r}_1^{[-]}(i)&=&\int_{|\hat{d}_i|<1} d\tilde\omega G_i(\tilde{\omega}) \sqrt{1-\left(\frac{\tilde\omega}{\sigma k_i \hat{R}_0}\right)^2},\nonumber \\
\eea
and  $\hat{c}_i$  and $\hat{d}_i$ indicating
\bea
\hat{c}_i=\frac{\omega-\Omega_0}{\sigma k_i \hat{R}_0R_1^{[-]}}, \quad \hat{d}_i=\frac{\tilde\omega}{\sigma k_i \hat{R}_0}.
\eea
Figure~\ref{fig:transitions} shows excellent agreement between the simulation results of Model 2 and the analytical prediction obtained in the annealed approximation for a Poisson network with average degree $c=12$ (panel (\textbf{a})) and for an uncorrelated  scale-free network with minimum degree $m=6$ and power-law exponent $\gamma=2.5$ (panel (\textbf{b})).  

Model 2 of explosive higher-order synchronization of coupled topological signals of nodes and links can be also solved exactly on a fully connected network.
Before discussing these theoretical results let us highlight that when treating Model 2 on a fully connected network the model parameters need to be rescaled appropriately to give a well defined transition in the large network limit.
In particular the coupling constant $\sigma$ and $\tau_1$ are rescaled to  
\bea
\sigma&\to &\frac{\sigma}{N},\nonumber \\
\tau_1 &\to &\tau_1\sqrt{N}. 
\eea
Moreover, for simplicity we set $\Omega_1=0$.
With these hypotheses the marginal distribution $G_i(\tilde{\omega})=G(\tilde{\omega})$  for every node $i$ of the network can be derived to be equal to 
\bea
G(\tilde{\omega})&=& \frac{\tau_1}{\sqrt{2\pi/\bar{c}}}\exp\left[-\tau_1^2\bar{c}\frac{\tilde{\omega}^2}{2}\right],
\label{marginal}
\eea
with $\bar{c}=N/(N-1)$.
The self-consistent equations for the order parameters $R_0=\hat{R}_0$ and $R_1^{[-]}=\hat{R}_1^{[-]}$ are given by 
\bea
&1 = \sigma   R_1^{[-]} h\left(\sigma^2  R_0^2 (R_1^{{[-]}})^2\right),\nonumber \\
&R_1^{[-]} = \sigma R_0 \tau_1 \sqrt{\bar{c}} h\left(\sigma^2 \tau_1^2 R_0^2 \right),\label{sys}
\eea
where the scaling function $h(x)$ is given by 
\bea
h(x) = \sqrt{\frac{\pi }{2}} e^{-x/4} \left[I_0\left(\frac{x}{4}\right)+I_1\left(\frac{x}{4}\right)\right],
\eea
with $I_0$ and $I_1$ indicating the modified Bessel functions.

These equations agree perfectly well with direct simulation of Model 2 dynamical Eqs. (\ref{model2}) on a fully connected network as it can be appreciated from Figure $\ref{fig:transitions}$(c). Moreover a closer look to these equations reveals an important aspect of these transitions.
While the backward transition has a well defined limit as $N\to \infty$, the forward transition occurs at larger value of the coupling constant for large network size $N$, and is only determined by finite size fluctuations, therefore the transition disappears in the limit $N\to\infty$.
Interestingly, this lack of a proper hysteresis loop can be also predicted for Model 2 defined on uncorrelated random graphs with finite second moment of the degree distribution, starting from their annealed approximation solution.

\section{Conclusions}

In this chapter our goal has been to provide evidence that the interplay between simplicial complex structure and dynamics is mediated by simplicial  geometry and topology. The spectral properties of the graph Laplacian and the higher-order Laplacian have been used here to reveal how simplicial synchronization is  shaped by topological and geometry of the simplicial complex.
In particular, we investigated how  simplicial network geometry changes the phase diagram of the  Kuramoto model defined on the network skeleton of simplicial complexes with notable geometrical properties and characterized by a  finite spectral dimension.
We have shown that a spectral dimension smaller or equal than four but larger than two can lead to a regime of frustrated synchronization characterized by large spatio-temporal fluctuations of the order parameter, while  a spectral dimension smaller or equal than two leads to a Kuramoto model in the incoherent state for every finite value of the coupling constant.
These theoretical results have been shown to apply to the  simplicial complexes generated by the modelling framework called Network Geometry with Flavor (NGF) that is able to generate simplicial complexes with tunable spectral dimension of the graph Laplacian. Interestingly, the NGF are characterized also by displaying higher-order spectral dimension of the higher-order up Laplacian that describe higher-order diffusion processes.

This chapter introduces also a set of models for capturing synchronization of topological signals, i.e., phases not only associated with the nodes of a simplicial complex but also to the higher-order simplices such as links, triangles, and so on.
This higher-order synchronization reveals itself in the order parameter of the irrotational and solenoidal projection of the topological signals. In the simple higher-order Kuramoto model the irrotational and solenoidal projection of the topological signal are uncoupled and undergo a sychronization transition at $\sigma_c=0$. However, when these two projections are coupled to each other by an adaptive global coupling the synchronization becomes explosive, i.e., discontinuous,  and occurs at a non-zero value of coupling constant. 

The higher-order Kuramoto model can be further extended to capture coupled topological signals of different dimension, for instance coupling phases associated with nodes and to links of a network or of a simplicial complex. This generalized higher-order Kuramoto model can lead to an explosive phase transition affecting simultaneously the phases associated with the nodes and the irrotational and solenoidal projection of the phases associated with the links. 
Interestingly, the higher-order Kuramoto model of coupled topological signals defined on nodes and links can be treated analytically using the annealed approximation when it is defined on a random uncorrelated network and can be solved exactly on a fully connected network. This solution confirms the discontinuous nature of the transition of the explosive higher-order Kuramoto model and sheds light on the stability of the hysteresis loop associated with the transition on finite networks. 
The mathematical framework that we have proposed here can be explored and modified in different directions and we believe that an in-depth analysis of the model and its variations will provide important insights on the interplay between topology and dynamics.  For instance, we note that the higher-order Kuramoto model has been recently modified \cite{lee} to investigate also the properties of a consensus model finding interesting results.

In conclusion, this chapter aims to provide an overview of the relation between network geometry topology and dynamics. We believe this topic will flourish in the incoming years and will transform our understanding of the relation between structure and dynamics of higher-order networks. Therefore our expectation is that this research line will play a relevant role for providing new insights in a variety of applications including  brain dynamics and biological transportation networks.

\bibliography{bib_Ana}
\bibliographystyle{unsrt}
\appendix
\section*{Appendix: Kuramoto dynamics expressed in terms of the incidence matrix}
In this Appendix our aim is to show that Eq. (\ref{KG}) that we rewrite here for convenience,
 \bea
\dot {\theta}_i=\omega_i+\sigma \sum_{j=1}^{N_{[0]}} a_{ij}\sin \left(\theta_j-\theta_i\right),
\label{KG2}
\eea
 is equivalent to Eq. (\ref{Kuramoto_B}) given by 
\bea
\dot{\bm{\theta}} = \bm{\omega} - \sigma \bm{B}_{[1]} \sin \bm{B}_{[1]}^T \bm{\theta}.
\label{Kuramoto_B2}
\eea
In order to show this let us observe that the incidence matrix ${\bf B}_{[1]}$ has elements given by 
\bea
[B_{[1]}]_{i\ell}=\left\{\begin{array}{cc} -1 &\mbox{if} \ \ell=[i,j],\\
1 &\mbox{if} \ \ell=[j,i],\\
0&\mbox{otherwise}.\end{array}\right.
\eea
To show the equivalence between Eq. (\ref{KG2}) and Eq. (\ref{Kuramoto_B2}) let us start by rewriting Eq. (\ref{Kuramoto_B2}) element by element, getting
\bea
\dot {\theta}_i=\omega_i-\sigma \sum_{\ell\in S_{1}} [B_{[1]}]_{i\ell}\sin \left(\sum_{j\in S_{0}}[B_{[1]}^{\top}]_{\ell j}\theta_j\right),
\label{Kuramoto_B3}
\eea
where  we indicate with $S_{1}$ the set of all links  present in the simplicial complex or network under consideration. 

Let us  consider the particular link $\ell=[i,j]$  in this case we have 
\bea
[B_{[1]}]_{i \ell}\sin \left(\sum_{j\in S_{0}}[B_{[1]}^{\top}]_{\ell j}\theta_j\right)=-a_{ij}\sin(\theta_j-\theta_i).
\eea
Equivalently, if we consider the same  link with opposite orientation $\ell=[j,i]$ we get
\bea
[B_{[1]}]_{i \ell}\sin \left(\sum_{j\in S_{0}}[B_{[1]}^{\top}]_{\ell j}\theta_j\right)=a_{ij}\sin(\theta_i-\theta_j)=-a_{ij}\sin(\theta_j-\theta_i).
\eea
Since the incidence matrix ${\bf B}_{[1]}$ has non-zero elements only among nodes and links incident to each other, it follows that  Eq. (\ref{KG2}) is equivalent to Eq. (\ref{Kuramoto_B2}).

\end{document}